\documentclass[manuscript]{aastex62}

\usepackage[figuresleft]{rotating}

\usepackage{placeins}
\usepackage{amsmath}
\usepackage{natbib}
\usepackage[normalem]{ulem}

\shorttitle{Suppression of Dielectronic Recombination II}
\shortauthors{Nikoli\'c et al.}

\begin{document}
\title{Suppression of Dielectronic Recombination Due to Finite Density Effects II: Analytical Refinement and Application to Density-dependent Ionization Balances and AGN Broad-line Emission}

\author{D.~Nikoli\'c}
\affil{Western Michigan University, Kalamazoo, MI, USA}

\author{T.~W.~Gorczyca}
\affil{Western Michigan University, Kalamazoo, MI, USA}

\author{K.~T. Korista}
\affil{Western Michigan University, Kalamazoo, MI, USA}

\author{M.~Chatzikos}
\affil{University of Kentucky, Lexington, KY, USA}

\author{G.~J.~Ferland}
\affil{University of Kentucky, Lexington, KY, USA}

\author{F.~Guzm\'{a}n}
\affil{University of Kentucky, Lexington, KY, USA}

\author{P~A.~M.~van~Hoof}
\affil{Royal Observatory of Belgium, Ringlaan 3, B-1180 Brussels, Belgium}

\author{R.~J.~R.~Williams}
\affil{AWE plc, Aldermaston, Reading RG7 4PR, UK}

\author{N.~R.~Badnell}
\affil{University of Strathclyde, Glasgow G4 0NG, UK}

\accepted{to The Astrophysical Journal Supplement Series}

\begin{abstract}
We present improved fits to our treatment of suppression of dielectronic recombination at intermediate densities.
At low densities, most recombined excited states eventually decay to the ground state, and therefore the total dielectronic recombination rate to all levels is preserved.
At intermediate densities, on the other hand, collisions can lead to ionization of higher-lying excited states, thereby suppressing the dielectronic recombination rate.
The improved suppression factors presented here, although highly approximate, allow summed recombination rate coefficients to be used to intermediate densities.
There have been several technical improvements to our previously presented fits.
For H- through B-like ions the activation log densities have been adjusted to better reproduce existing data.
For B-, C-, Al-, and Si-like ions secondary autoionization is now included.
The treatment of density discontinuity in electron excitations out of ground state H-, He-, and Ne-like ions has been improved.
These refined dielectronic recombination suppression factors are used in the most recent version of the plasma simulation code Cloudy.
We show how the ionization and emission spectrum change when this physics is included.
Although these suppression factors improve the treatment of intermediate densities, they are highly approximate and are not a substitution for a complete collisional-radiative model of the ionization balance.
\end{abstract}


%===============================%
%===% SECTION: INTRODUCTION %===%
%===============================
\section{Introduction}
\label{intro}

Dielectronic Recombination (DR) is an important process that determines the ionization balance in cosmic plasmas.
To this end, a large effort has been devoted to computing a reliable database for total and partial DR rate coefficients \cite[see][and the 14 subsequent papers in that series, as referenced by the latest one, \cite{Kaur:2017}]{drproject}.
These data are necessary input to plasma simulation codes such as Cloudy \citep{cloudy17}.
However, all of that data have been computed assuming a zero-density plasma environment, reducing the total DR problem to a more tractable atomic physics problem consisting of a single incoming electron colliding with a single atomic ion and recombining to an ionization state one charge lower, with the emission of one photon (and any additional, cascading photons).

It has long been recognized \citep{burgsum} that in a plasma of non-negligible density, such as in the broad emission-line regions of Active Galactic Nuclei (AGNs), with densities $n_{\rm{e}} \sim 10^{10}\ {\rm cm}^{-3}$, additional, secondary plasma electrons enter into the problem and may affect the total recombination rate via intermediate electron-impact ionization of captured, doubly excited resonance states, depleting the radiative rate and thereby the final recombination probability.
Treating this more complex problem requires, in addition to accurate, zero-density atomic data, a generalized collisional-radiative (GCR) model approach \citep{sumhoop} to account for all possible recombination and ionization pathways.

To date, there has been limited GCR modeling carried out, and we have relied on the pioneering work of \citet{burgsum}, and the extensive, detailed calculations for the density, temperature, and elemental-dependent, effective  recombination rate coefficient of \citet{summersRAL}, as a guide for quantifying the suppression of DR due to finite density effects.
This was the approach we adopted in a previous publication \cite[hereafter referred to as Paper~I]{Nikolic:2013}.

After several model applications of this algorithm, it was found in certain situations (see, for example, \cite{Young:2018}), that the original formulation was susceptible on finer grids to numerical difficulties arising from a discontinuity in temperature of the effective DR rate coefficient.
This problem affects the first five isoelectronic sequences: H-like through B-like.

The present paper serves three purposes.
First, a minor ``tweak'' to our previous formulation is introduced to circumvent the earlier discontinuity in temperature of the suppression factor.
The second goal is to provide an alternative suppression factor for four sequences, following \cite{summersRAL}, depending on the source of (physics included in) the zero-density DR rate coefficients the factor is to be applied to.
Third, representative finite-density plasma simulations are carried out using the new, modified Cloudy version to assess the effect of finite densities, via the consequent DR suppression, in an actual plasma environment.

\section{Generalized Density Suppression Model}
\label{Sec:Suppress}

The present approach for treating DR suppression closely follows the original formulation of \cite{Nikolic:2013}, with only minor refinement in the final algorithm, but for completeness and to avoid any confusion, the entire formulation is repeated below, with the important modification highlighted.
In general, the {\em effective} DR rate coefficient $\alpha^{\rm{eff}}_{\rm{DR}}(n_{\rm{e}},T,q,N)$, as a function of electron density $n_{\rm{e}}\,(\rm{cm}^{-3})$ and temperature $T \,(\rm{K})$, ionic charge state $q$, and isoelectronic sequence (labeled by $N$), is suppressed from the zero-density value $\alpha_{\rm{DR}}(T)\,(\rm{cm}^{3}\rm{s}^{-1})$ by a dimensionless suppression factor $S^N( n_{ \rm{e} },T;q )$,
\begin{eqnarray}
\alpha^{\rm{eff}}_{\rm{DR}}(n_{\rm{e}},T,q,N) & \equiv & S^N(n_{\rm{e}},T;q)\alpha_{\rm{DR}}(T)\ \ ;
\label{eqsuppress}
\end{eqnarray}
for simplicity, we use the dimensionless log density parameter $x=\log_{10}n_{\rm{e}}$.

The functional form of $S^N(n_{\rm{e}},T,q)$ is taken to be a pseudo-Voigt profile
 \begin{equation}
S^N(x,T;q) = \left\{
 \begin{matrix}
 1  &  x \le  x_a(T;q,N)\\
 {\rm e}^{-(\frac{x-x_a(T;q,N)}{w/\sqrt{\ln 2}})^{2}}  &  x \geq x_a(T;q,N)
 \end{matrix}\right. \ ,
 \label{eqsuppression2}
\end{equation}
of width $w=5.64586$ and an activation log density $x_a(T;q,N)$ that is represented by the complicated expression
\begin{eqnarray}
x_a(T;q,N) & = & x_{a}^{0} + log_{10}\left[\left(\frac{q}{q_0(q,N)}\right)^{7}\left(\frac{T}{T_0(q,N)}\right)^{1/2}\right]\ . \label{eqxanew}
\end{eqnarray}
A fit of the suppression factors of \citet{summersRAL} for all ions yielded a global (log) activation density $x_{a}^{0}=10.1821$ and more complicated expressions for the zero-point temperature $T_0 \,(\rm{K})$ and charge state $q_0$.
These were found to depend on both the  ionic charge state $q$ and the isoelectronic sequence $N$ viz.
\begin{eqnarray}
T_0(q,N) &= &5\times10^{4}\,[q_0(q,N)]^{2} \label{t0}
\end{eqnarray}
and
\begin{eqnarray}
q_0(q,N) & = &  (1 - \sqrt{2/3q})A(N)/\sqrt{q}\ , \label{q0}
\end{eqnarray}
where
\begin{eqnarray}
A(N) & = &  12 + 10N_{1} + \frac{10N_{1}-2N_{2}}{N_{1}-N_{2}}(N-N_{1}) \label{AN}
\end{eqnarray}
depends on the isoelectronic sequence in the periodic table according to the specification of the parameters
\begin{eqnarray}
(N_{1},N_{2}) & = &
\begin{pmatrix}
  (3,10)  & N\in 2^{nd} \, \rm{row}  && (37,54)  & N\in 5^{th} \, \rm{row} \\
  (11,18) & N\in 3^{rd} \, \rm{row}  && (55,86)  & N\in 6^{th} \, \rm{row} \\
  (19,36) & N\in 4^{th} \, \rm{row}  && (87,118) & N\in 7^{th} \, \rm{row} \\
\end{pmatrix}\ .
\label{N1N2}
\end{eqnarray}
If the zero-density DR data $\alpha_{\rm{DR}}(T)$ in Eq.~(\ref{eqsuppress}) neglects the secondary autoionization \citep{Blaha:1972}, this parameterization is sufficient for all isoelectronic sequences $N\ge 6$.
However, the given parameterization was not flexible enough to provide an adequate fit to the \citet{summersRAL} data for the lower isoelectronic sequences $N\le 5$.
Instead, we explicitly list the optimal values for $A(N)$, for lower ionization stages, in Table~\ref{TableAN}.

Even with this formulation, an additional  modification was necessary at electron temperatures and/or ionic charges for which the $q$-scaled temperature $\theta \equiv T/q^2$  was very low ($\theta\leq 2.5\times10^{4}$ K), which is now a slightly different formulation than that used previously.

In Paper I \cite{Nikolic:2013}, we modified the factor $A(N)$ for low temperatures as follows:
\begin{eqnarray}
A^{mod,old}(N\le 5)  = & \left\{
\begin{array}{rr}
 A(N)\ , &  \theta > 2.5\times10^{4}\ {\rm K}\\
 2\times A(N)\ , & \theta \le 2.5\times10^{4}\ {\rm K}
\end{array}
\right. \ .
\label{eqanold}
\end{eqnarray}
Using this algorithm, the discontinuity in the modification factor, from unity to a factor of two at $\theta=T/q^2=2.5\times 10^4$ K, was found to cause numerical difficulties for certain density-dependent modeling applications, using the previous Cloudy release following Paper~I.
In order to avoid any such algorithmic difficulties in the future, and also to allow for an improved fit of the available suppression factor data of \citet{summersRAL} by a generalized suppression formulation, we update the additional low-temperature modification factor via a continuous function:
\begin{eqnarray}
A^{mod}(N)  = & \left\{
\begin{array}{lr}
\psi^N(q,T) \times A(N), & N\le 5 \\
\psi^{N}_{sec}(q,T) \times A(N), & \;N = 5, 6, 13, 14 \\
\end{array}
\right. \ .
\label{eqannew}
\end{eqnarray}
Here, the additional dimensionless functions,
 \begin{eqnarray}\label{eqFactor}
\psi^N(q,T) & = & 2\times \frac{1 + \pi_{3} \times {\rm e}^{-(\frac{\log_{10}T-\pi_{1}}{\sqrt{2}\pi_{2}})^{2}} + \pi_{6} \times {\rm e}^{-(\frac{\log_{10}T-\pi_{4}}{\sqrt{2}\pi_{5}})^{2}}}{1 + {\rm e}^{-\sqrt{25000q^2/T}}} \\
\pi_{i} & = & \pi_{i}^{(1)} + \pi_{i}^{(2)} \times q^{\pi_{i}^{(3)}} \times {\rm e}^{- q / \pi_{i}^{(4)} } \; i = 1 \ldots 6, \nonumber
\end{eqnarray}
\begin{eqnarray}\label{eqFactorSec}
\psi^{N}_{sec}(q,T) & = & 1 + \gamma_{3} \times {\rm e}^{-(\frac{\log_{10}T-\gamma_{1}}{\sqrt{2}\gamma_{2}})^{2}} + \gamma_{6} \times {\rm e}^{-(\frac{\log_{10}T-\gamma_{4}}{\sqrt{2}\gamma_{5}})^{2}} \\
\gamma_{i} & = & \gamma_{i}^{(1)} + \gamma_{i}^{(2)} \times q^{\gamma_{i}^{(3)}} \times {\rm e}^{- q / \gamma_{i}^{(4)} } \; i = 1 \ldots 6, \nonumber
\end{eqnarray}

are continuous at all temperatures and ensure the same asymptotic behavior as determined before,
\begin{eqnarray}
A^{mod}(N) & {\longrightarrow \atop \theta\rightarrow \infty } &  A(N) \\
& {\longrightarrow \atop \theta\rightarrow 0 } & 2\times A(N) \ ,
\label{eqanasymp}
\end{eqnarray}
and the additional flexibility introduced allows for an improved fit to the \citet{summersRAL} data;
the adjustment coefficients $\pi_{i}^{(j)}$ and $\gamma_{i}^{(j)}$ are given in Table~\ref{TablePI} and the $\psi^N(q,T)$ for iso-electronic sequences considered here are illustrated in Fig.~\ref{Appendix:figpsi} of Appendix~\ref{Appendix:Psi}.

If the main concern is to remove the temperature discontinuity while keeping the overall agreement with \citet{summersRAL} data to better than 25\%, then we suggest using the ``simplified" part of Table~\ref{TablePI}.
However, for an overall agreement with \citet{summersRAL} data of 14\% and better, the use of the ``detailed" part of Table~\ref{TablePI} is recommended for five lowest isoelectronic sequences.
For the B-, C-, Al-, and Si-like sequences, the effects of secondary autoionization cannot be neglected.
If the zero-density DR data $\alpha_{\rm{DR}}(T)$ in Eq.~(\ref{eqsuppress}) for these isoelectronic sequences already account for secondary autoionization effects, then the ``secondary autoionization" part of Table~\ref{TablePI} should be used.
Note that Table~\ref{TablePI} contains two sets of adjustment coefficients for B-like ions, depending on whether the zero-density DR data $\alpha_{\rm{DR}}(T)$ in Eq.~(\ref{eqsuppress}) already contain corrections due to secondary autoionization or not.
The results of Paper~I should be used for all other isoelectronic sequences, including C-, Al-, and Si-like sequences if being applied to zero-density DR rate coefficients, which do not include secondary autoionization.
In the 2017 release of Cloudy \citep{cloudy17} the zero-density DR data for B-like ions is modified using the ``secondary autoionization" part of Table~\ref{TablePI} in accordance with modern DR data of \cite{drproject}\footnote{H- through Si-like data is available from \url{http://amdpp.phys.strath.ac.uk/tamoc/DATA/DR/}}, which include the effect.
For details regarding the variation of accuracy with respect to approximations used over a wide range of temperatures and isoelectronic sequences, see Fig.~\ref{Appendix:figsurveyOldSmoothGreen} of Appendix~\ref{Appendix:Acuracy}.

\begin{deluxetable}{rccrcc}
%\tabletypesize{\tiny}
\tabletypesize{\small}
\tablecaption{Modified $A(N)$ coefficients from Eq.~(\ref{AN}).
\label{TableAN}}
\tablewidth{0pt}
\tablehead{
\colhead{Sequence} & \colhead{ $N$} & \colhead{$A(N)^{\dag}$} & \colhead{Sequence}  & \colhead{ $N$} & \colhead{$A(N)^{\ddag}$}
}
\startdata
\multicolumn{3}{c}{No Secondary Autoionization} & \multicolumn{3}{c}{Secondary Autoionization Included} \\
 H-like & 1 & 16 &  B-like & 5 & 52  \\
He-like & 2 & 18 &  C-like & 6 & 37.7\\
Li-like & 3 & 66 & Al-like & 13& 100.9  \\
Be-like & 4 & 66 & Si-like & 14& 90.3\\
 B-like & 5 & 52 &&&  \\
\hline
\multicolumn{6}{l}{ ${}^{\dag}$ these must be multiplied by $\psi^N(q,T)$ given in Eq.~(\ref{eqFactor}) } \\
\multicolumn{6}{l}{ ${}^{\ddag}$ these must be multiplied by $\psi^{N}_{sec}(q,T)$ given in Eq.~(\ref{eqFactorSec}) } \\
\hline\enddata
\end{deluxetable}

%\pagebreak
%\vskip -0.5in
\begin{deluxetable}{rcc|cccc|cccc||rcc|cccc}
%\tabletypesize{\scriptsize}
%\tabletypesize{\small}
\tabletypesize{\tiny}
\tablecaption{Adjustment Coefficients $\pi_{i}^{(j)}$ from Eq.~(\ref{eqFactor}) and $\gamma_{i}^{(j)}$ from Eq.~(\ref{eqFactorSec}).
\label{TablePI}}
\tablewidth{0pt}
\tablehead{
\hline\hline
\multicolumn{3}{r}{Adjustment Factor:} &
\multicolumn{4}{c}{``Detailed" $\psi^N(q,T)$} &
\multicolumn{4}{c}{``Simplified" $\psi$} &
\multicolumn{7}{c}{``Secondary Autoionization" $\psi^{N}_{sec}(q,T)$} \\
\colhead{Sequence}          &
\colhead{ $N$}              &
\colhead{$\pi_{i}$}         &
\colhead{$\pi_{i}^{(1)}$}   &
\colhead{$\pi_{i}^{(2)}$}   &
\colhead{$\pi_{i}^{(3)}$}   &
\colhead{$\pi_{i}^{(4)}$}   &
\colhead{$\pi_{i}^{(1)}$}   &
\colhead{$\pi_{i}^{(2)}$}   &
\colhead{$\pi_{i}^{(3)}$}   &
\colhead{$\pi_{i}^{(4)}$}   &
\colhead{Sequence}          &
\colhead{ $N$}              &
\colhead{$\gamma_{i}$}      &
\colhead{$\gamma_{i}^{(1)}$}&
\colhead{$\gamma_{i}^{(2)}$}&
\colhead{$\gamma_{i}^{(3)}$}&
\colhead{$\gamma_{i}^{(4)}$}
} % end of tablehead
\startdata %done
%y0 + a * q^b * exp(-q/c)y0  a  b c y0abc
H-like &  1 & $\pi_{1}$  &  4.7902  & 0.32456 &  0.97838 & 24.78084 & 0        & 0 & 0 & $\infty$ &C-like &  6 & $\gamma_{1}$  &5.90184  &-1.2997   &  1.32018 &  2.10442   \\ % xc1
       &    & $\pi_{2}$  & -0.0327  & 0.13265 &  0.29226 & $\infty$ & $\infty$ & 0 & 0 & $\infty$ &       &    & $\gamma_{2}$  &0.12606  &0.009     &  8.33887 &  0.44742   \\ % w1
       &    & $\pi_{3}$  & -0.66855 & 0.28711 &  0.29083 &  6.65275 & 0        & 0 & 0 & $\infty$ &       &    & $\gamma_{3}$  &-0.28222 &0.018     &  2.50307 &  3.83303   \\ % A1
       &    & $\pi_{4}$  &  6.23776 & 0.11389 &  1.24036 & 25.79559 & 0        & 0 & 0 & $\infty$ &       &    & $\gamma_{4}$  &6.96615  &-0.41775  &  2.75045 &  1.32394   \\ % xc2
       &    & $\pi_{5}$  &  0.33302 & 0.00654 &  5.67945 &  0.92602 & $\infty$ & 0 & 0 & $\infty$ &       &    & $\gamma_{5}$  &0.55843  &0.45      &  0.0     &  2.06664   \\ % w2
       &    & $\pi_{6}$  & -0.75788 & 1.75669 & -0.63105 &184.82361 & 0        & 0 & 0 & $\infty$ &       &    & $\gamma_{6}$  &-0.17208 &-0.17353  &  0.0     &  2.57406   \\ % A2
\hline % done
He-like&  2 & $\pi_{1}$  &  4.82857 & 0.3     &  1.04558 & 19.6508  & 0        & 0 & 0 & $\infty$ &Al-like& 13 & $\gamma_{1}$  &6.59628  &-3.03115  &  0.0     &  10.519821 \\ % xc1
       &    & $\pi_{2}$  & -0.50889 & 0.6     &  0.17187 & 47.19496 & $\infty$ & 0 & 0 & $\infty$ &       &    & $\gamma_{2}$  &1.20824  &-0.85509  &  0.21258 &  25.56     \\ % w1
       &    & $\pi_{3}$  & -1.03044 & 0.35    &  0.3586  & 39.4083  & 0        & 0 & 0 & $\infty$ &       &    & $\gamma_{3}$  &-0.34292 &-0.06013  &  4.09344 &  0.90604   \\ % A1
       &    & $\pi_{4}$  &  6.14046 & 0.15    &  1.46561 & 10.17565 & 0        & 0 & 0 & $\infty$ &       &    & $\gamma_{4}$  &7.92025  &-3.38912  &  0.0     &  10.02741  \\ % xc2
       &    & $\pi_{5}$  &  0.08316 & 0.08    &  1.37478 &  8.54111 & $\infty$ & 0 & 0 & $\infty$ &       &    & $\gamma_{5}$  &0.06976  &0.6453    &  0.24827 &  20.94907  \\ % w2
       &    & $\pi_{6}$  & -0.19804 & 0.4     &  0.74012 &  2.54024 & 0        & 0 & 0 & $\infty$ &       &    & $\gamma_{6}$  &-0.34108 &-0.17353  &  0.0     &  6.0384    \\ % A2
\hline %done
Li-like&  3 & $\pi_{1}$  & 4.55441  & 0.08    &  1.11864 &  $\infty$& 0        & 0 & 0 & $\infty$ &Si-like& 14 & $\gamma_{1}$  &5.54172  &-1.54639  &  0.01056 &  3.24604   \\ % xc1
       &    & $\pi_{2}$  & 0.3      & 2.0     & -2.0     &  67.36368& $\infty$ & 0 & 0 & $\infty$ &       &    & $\gamma_{2}$  &0.39649  &0.8       &  3.19571 &  0.642068  \\ % w1
       &    & $\pi_{3}$  & -0.4     & 0.38    &  1.62248 &  2.78841 & 0        & 0 & 0 & $\infty$ &       &    & $\gamma_{3}$  &-0.35475 &-0.08912  &  3.55401 &  0.73491   \\ % A1
       &    & $\pi_{4}$  & 4.00192  & 0.58    &  0.93519 &  21.28094& 0        & 0 & 0 & $\infty$ &       &    & $\gamma_{4}$  &6.88765  &-1.93088  &  0.23469 &  3.23495   \\ % xc2
       &    & $\pi_{5}$  & 0.00198  & 0.32    &  0.84436 &  9.73494 & $\infty$ & 0 & 0 & $\infty$ &       &    & $\gamma_{5}$  &0.58577  &-0.31007  &  3.30137 &  0.83096   \\ % w2
       &    & $\pi_{6}$  & 0.55031  & -0.32251&  0.75493 &  19.89169& 0        & 0 & 0 & $\infty$ &       &    & $\gamma_{6}$  &-0.14762 &-0.16941  &  0.0     &  18.53007  \\ % A2
\hline % done
Be-like&  4 & $\pi_{1}$  & 2.79861  & 1.0     &  0.82983 &18.05422  & 0        & 0 & 0 & $\infty$ &       &    &               &         &          &          &            \\
       &    & $\pi_{2}$  & -0.01897 & 0.05    &  1.34569 &10.82096  & $\infty$ & 0 & 0 & $\infty$ &       &    &               &         &          &          &            \\
       &    & $\pi_{3}$  & -0.56934 & 0.68    &  0.78839 & 2.77582  & 0        & 0 & 0 & $\infty$ &       &    &               &         &          &          &            \\
       &    & $\pi_{4}$  & 4.07101  & 1.0     &  0.7175  &25.89966  & 0        & 0 & 0 & $\infty$ &       &    &               &         &          &          &            \\
       &    & $\pi_{5}$  & 0.44352  & 0.05    &  3.54877 & 0.94416  & $\infty$ & 0 & 0 & $\infty$ &       &    &               &         &          &          &            \\
       &    & $\pi_{6}$  & -0.57838 & 0.68    &  0.08484 & 6.70076  & 0        & 0 & 0 & $\infty$ &       &    &               &         &          &          &            \\
\hline % done
B-like &  5 & $\pi_{1}$  &6.75706   &-3.77435 &  0.0     & 4.59785  & 0        & 0 & 0 & $\infty$ &B-like &  5 & $\gamma_{1}$  &6.91078  &-1.6385   &  2.18197 &  1.45091   \\ % xc1
       &    & $\pi_{2}$  &0.0       &0.08     &1.34923   & 7.36394  & $\infty$ & 0 & 0 & $\infty$ &       &    & $\gamma_{2}$  &0.4959   &-0.08348  &  1.24745 &  8.55397   \\ % w1
       &    & $\pi_{3}$  &  -0.63   &0.06     &  2.65736 & 2.11946  & 0        & 0 & 0 & $\infty$ &       &    & $\gamma_{3}$  &-0.27525 &0.132     &  1.15443 &  3.79949   \\ % A1
       &    & $\pi_{4}$  &7.74115   &-4.82142 &  0.0     & 4.04344  & 0        & 0 & 0 & $\infty$ &       &    & $\gamma_{4}$  &7.45975  &-2.6722   &  1.7423  &  1.19649   \\ % xc2
       &    & $\pi_{5}$  &0.26595   &0.09     &  1.29301 & 6.81342  & $\infty$ & 0 & 0 & $\infty$ &       &    & $\gamma_{5}$  &0.51285  &-0.60987  &  5.15431 &  0.49095   \\ % w2
       &    & $\pi_{6}$  &-0.39209  &0.07     &  2.27233 & 1.9958   & 0        & 0 & 0 & $\infty$ &       &    & $\gamma_{6}$  &-0.24818 &0.125     &  0.59971 &  8.34052   \\ % A2
\hline
\enddata
\end{deluxetable}

\begin{figure}[!hbtp]
\centering
\includegraphics[width=0.7\columnwidth]{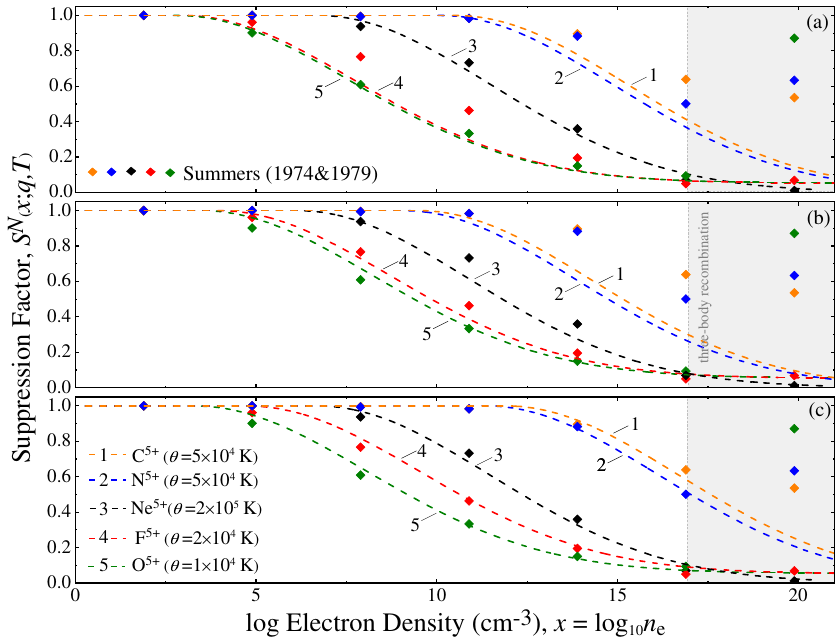}
\caption{Computed suppression factors for representative situations (ions, sequences, temperatures, and densities) as compared to the GCR results of \citet{summersRAL}.
 Results correspond to cases when the activating log density $x_a(T;q,N)$ is estimated using the earlier formulation of Paper~I (a), the ``simplified" $\psi$ (b), or the ``detailed" $\psi^N(q,T)$ (c) using the adjustment factors given in Eq.~(\ref{eqFactor}) and Table~\ref{TablePI}.
\label{figsgcr}
}
\end{figure}

To illustrate how much better the present algorithm reproduces the \citet{summersRAL} suppression factor, we show a comparison of old and new results in Fig.~\ref{figsgcr} for several representative ions, sequences, and temperatures as a function of electron density.

For even lower temperatures, we add a final modification to ensure that, at plasma energies $kT$ much less than the excitation energies, $\epsilon_N(q)$, for which the intermediate resonance states are not suppressed (see Paper~I), the suppression is ``turned off'':
\begin{eqnarray}
S^N(x,T;q) & = & 1 - \left[ 1 - S^{N}(x,T;q) \right]\times \exp\left(-\frac{\epsilon_N(q)}{10kT}\right) \ .
\label{eqsmod}
\end{eqnarray}
When compared to the Paper~I methodology for H-, He-, and Ne-like ions, in the present study we ``turn off'' the suppression for these ions in continuous fashion with respect to the global activation log density $x_{a}^{0}$; see Table~\ref{Appendix:TableEnergies} of Appendix~\ref{Appendix:ExcitationEnergies}.
We also update the excitation energy $\epsilon_{14}(2)$ for $S^{2+}$ following the results of \citet{Badnell:2015}.
The excitation energies for other isoelectronic sequences remain the same as in Paper~I, parameterized by the expression
\begin{eqnarray}
\epsilon_N(q) & = & \sum_{j=0}^5 p_{N,j} \left(\frac{q}{10}\right)^j\ .
\label{eqepsilon}
\end{eqnarray}
As in Paper~I, these parameters are optimized using the available NIST excitation energies \citep{nist} and are listed in Table~\ref{Appendix:TableEnergies} of Appendix~\ref{Appendix:ExcitationEnergies}.

\section{Ionization and emission predictions}
The density dependence of the ionization rate coefficient at most astrophysical densities is negligible compared to that of the (dielectronic) recombination one -- see e.g. Sec.~3.2 of \citet{summersRAL}.
This is a reasonable approximation since the initial state population for ionization is almost exclusively in the ground states (and perhaps metastables), which have little density dependence, compared to the excited states.
In contrast, density dependence in recombination arises via the final state, and in DR these are highly excited.
The effective "density-dependent" ionization rate coefficients can be downloaded from Open ADAS \citep{OPEN-ADAS} in ADF11 data format at two degrees of refinement: (i) ``unresolved,'' in which ions are assumed to be in the ground state only, and (ii) ``metastable-resolved,'' in which both ground and metastable states of ions may be dominant.
Appendix~\ref{Appendix:CloudyExamples} presents the ionization balance for the thirty lightest elements for the photoionization and collisional ionization cases.

\subsection{The equivalent two-level approximation}
Several approaches can be taken for computing the ionization distribution of the elements.
In the equivalent two-level approximation, which applies at low densities, recombinations to excited states will eventually decay to the ground state.
Only ionizations from the ground state need to be considered, since at low densities this is where nearly all of the population lies.
This approximation holds for the interstellar medium (ISM) and is described in texts such as \cite[hereafter AGN3]{Osterbrock:2006}, and in Section~3.2 of \citep[hereafter C17]{cloudy17}.
In this approximation, summed recombination coefficients, such as those given at \url{http://amdpp.phys.strath.ac.uk/tamoc/DATA}, can be used.
At high densities, the gas comes into LTE and the ionization is given by the Saha-Boltzmann equation.
This limit is reached in the lower parts of many stellar atmospheres and accretion disks \citep{Hubeny:2014}.
The intermediate-density case is the most difficult since neither limit applies and collisional processes affecting the highly excited Rydberg levels must be taken into account.
In this case a ``collisional-radiative model" (CRM) must be used.
Such models are discussed in \cite{Ralchenko:2016} and Section 3.1 of C17.
Section~3 of C17 used Cloudy's full CRM treatment of one- and two-electron systems to make estimates of the range over which the two-level and LTE approximations hold.
The ranges are significantly different for collisionally and photoionized environments.
CRM effects are important at much lower densities in the collisional case due to the dominance of near-threshold collisional ionization, which also affect the Rydberg level populations.
In the photoionized case, the gas kinetic temperature is much lower than the ionization potentials so collisional ionization is much less important.  
The range over which the two-level approximation works is also very strongly density-dependent.
The two-level approximation works at much higher densities for higher charges $q$ due to the well-known $q^{-7}$ scaling of collisional effects, described by \cite{bates} and \cite{burgsum}.
This paper develops corrections to the summed recombination coefficients to improve the behavior of the two-level approximation at intermediate densities.
The results of this paper are included in the C17.01 update to Cloudy and we use that version in the calculations presented here.

\subsection{The case of Oxygen}
\begin{figure}[!htbp]
\centering
\includegraphics[width=0.7\columnwidth]{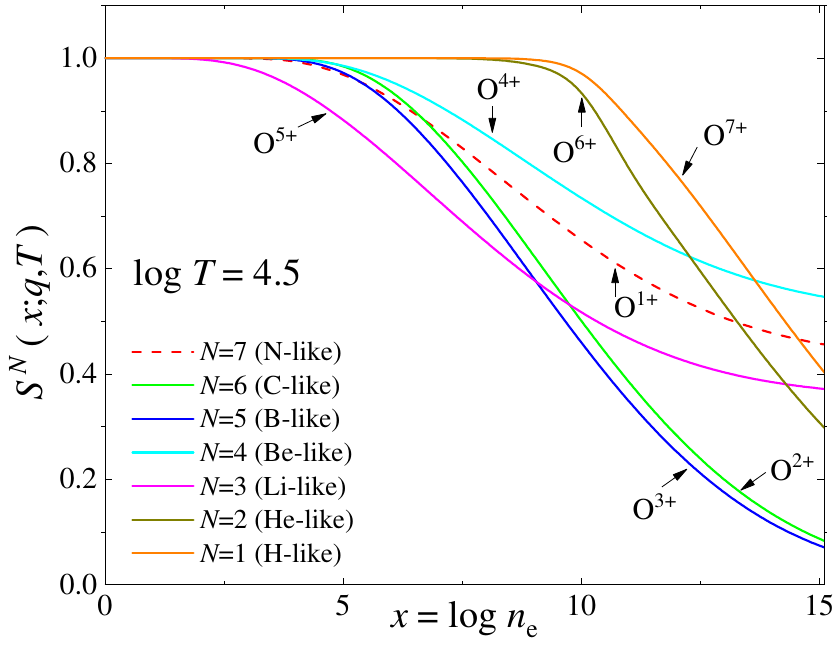}
\caption{Suppression of oxygen DR for various ions and a temperature of 10$^{4.5}$~K. The legend indicates the isoelectronic sequence of the recombining species, with O$^{1+}$ indicating recombination forming O~I or O$^{0}$.
The logarithm of electron density is indicated along the independent axis.
\label{Oxygen}
}
\end{figure}

We first focus on oxygen since it is the third most common element, has high quality DR rates, and produces strong emission lines from the IR to the X-ray so has great astronomical importance.
Figure~\ref{Oxygen} shows the suppression factors for the first seven ionization stages of oxygen.  
These were computed for a gas kinetic temperature of 10$^{4.5}$~K and various electron densities, indicated along the independent axis.  
This low temperature is characteristic of photoionized plasmas with a moderate level of ionization and is chosen to illustrate the physics.

The density and charge dependencies reflect the decays of the highly excited levels.
Suppression is negligible for densities below $\sim 10^4$~cm$^{-3}$.
For very low densities, the collisional rate is much slower than the radiative decay rates, so electrons captured into Rydberg levels will undergo a stabilizing radiative decay and the ion recombines.
The detailed density dependence is different for different ions because the electron configuration affects the detailed stabilization channels, but the tendency is for the importance of suppression to decrease with increasing ionization, a tendency also shown for the one- and two-electron species in Section~3 of C17.
The radiative decay rates, which stabilize the recombined ion, have a rapid charge dependence, $\sim q^4$, while the collisional ionization rate coefficients decrease.
So, for higher charges $q$, higher densities are needed to obtain the same suppression effect, according to the $\sim q^{-7}$ effect discussed by \cite{bates} and \cite{burgsum}.
The remainder of this section develops collisional- and photoionized models with and without this suppression to study its effects on spectroscopic models.
We note that \cite{summersRAL} did not provide any finite-density data for the recombination of singly charged ions to form neutrals.
Consequently, results for neutrals should be treated with extreme caution since they follow from extrapolation of doubly to singly charged data.

\subsection{Ionization calculations for the 30 lightest elements}
We consider two classes of models: a model in electron collisional ionization equilibrium, and one for a photoionized gas.
We note that a significant amount of C, O, Si, and S form molecules in the lowest temperature and electron density collisional model.
Although physically correct, this introduces a distraction from our main point, the density-dependent effect of DR suppression upon the ionization.
The chemistry network was disabled for the calculations presented here, which has the added benefit of decoupling the results from uncertainties in the chemical rates and the completeness of the chemical database.
We concentrate now on oxygen and show our results in Figure~\ref{Ionization} with corresponding functions $\psi^N(q,T)$ illustrated in Figure~\ref{Appendix:figpsi} of Appendix~\ref{Appendix:Psi}.
\begin{figure}[!htbp]
\centering
\begin{tabular}{l}
\includegraphics[width=0.7\textwidth]{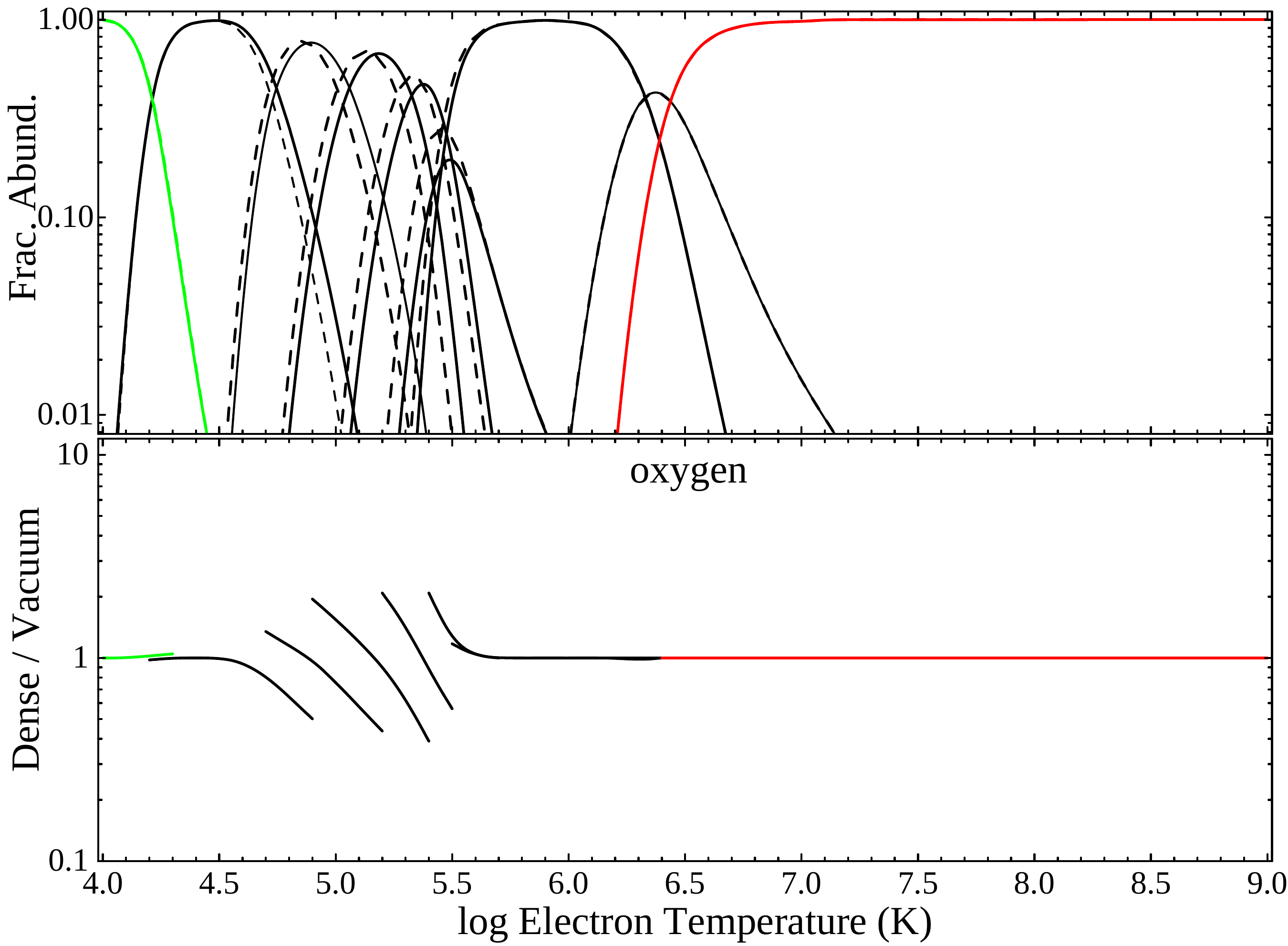} \\
\includegraphics[width=0.7\textwidth]{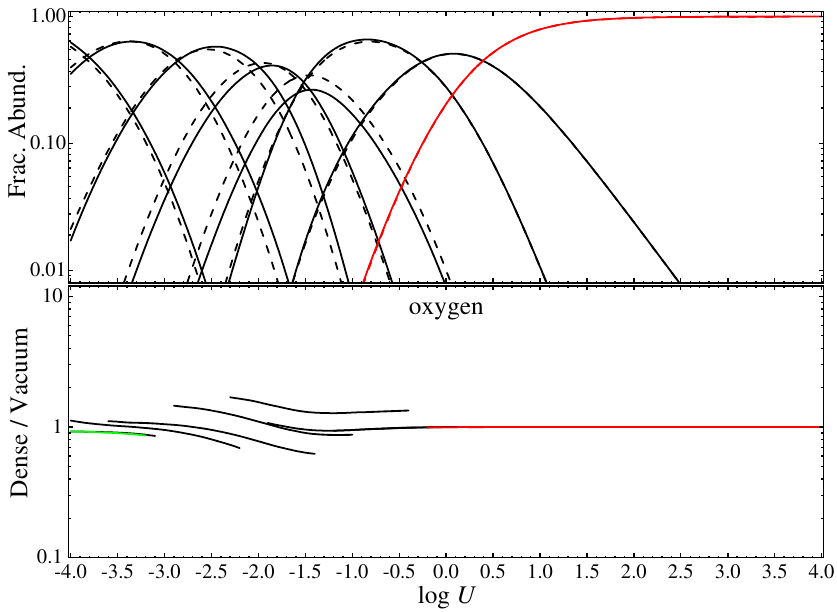}
\end{tabular}
\caption{Ionization results for oxygen for the two cases.
Appendix~\ref{Appendix:CloudyExamples} shows similar results for the 30 lightest elements.
The upper pair of panels is for a collisionally ionized gas and the independent axis is the gas temperature.
The lower pair of panels is for photoionization and the ionization parameter is the independent axis.
In each, the upper sub-panel shows the ionization at densities of 1~cm$^{-3}$ (vacuum, solid line) and 10$^{10}$~cm$^{-3}$ (dense, dashed line), while the lower sub-panel shows their ratio.
\label{Ionization}
}
\end{figure}

In the electron collisional case, ionizing photons can be neglected and only impact ionization by thermal electrons is important.
Ionizations by other particles such as protons and helium nuclei are included but are generally negligible.
As shown in the discussion around Equation~(4) of C17, the ionization fraction has no direct density dependence if the collisional ionization and recombination rate coefficients do not depend on density.
The ionization fraction depends only on the temperature due to the exponential dependence of the Boltzmann factor in the collisional ionization rate coefficient and the slower temperature dependencies of the recombination rate coefficients.  
Higher ionization is produced by higher temperatures, the free parameter in this case.
The temperature is varied over a very wide range so that the charge states of most elements range from fully atomic at low temperatures to bare nuclei at high values.

We also consider photoionized clouds.
Here, the radiation field is the dominant source of ionization.
The photoionization rate has no temperature dependence, so the recombination rate coefficients introduce the only direct temperature dependence.
That temperature is determined by the balance between heating and cooling processes, as discussed in Chapter~3 of AGN3.
Increases in the ionization are produced by either a brighter radiation field, which increases the photoionization rate, or by a smaller electron density, which decreases the recombination rate.
The ionization parameter $U$, the dimensionless ratio of photon to hydrogen densities, (AGN3, Equation~(14.7)), is defined as
\begin{equation}
U \equiv \frac{Q({\rm H})}{4 \pi r^{2} n({\rm H})\;c} \equiv \frac{\Phi({\rm H})}{n({\rm H})\;c}\ ,
\label{eqepsilon}
\end{equation}
where $Q({\rm H})$ is the total number of ionizing photons, $r$ is the separation between the radiation source and the cloud, and $\Phi({\rm H})$ is the flux of hydrogen-ionizing photons, $n({\rm H})$ is the number density of hydrogen, and $c$ is the speed of light.
This parameter plays the same role as the temperature in the collisional case.
We vary $U$ over a broad range to change the ionization from atomic to fully ionized.
The gas is irradiated by a continuum with $f_{\nu} \propto \nu^{-1}$ between 30 meV and 100 MeV.

In photoionization equilibrium, the gas temperature depends on the ionization parameter in a complex way, but generally tends to increase with $U$, and at constant $U$ it increases with density due to suppression of collisional cooling at high densities.
These temperature changes would obfuscate the central point of this paper, the density-dependent suppression, since we wish to compare models with different densities, which will have different temperatures.
To remove this confusion, we artificially set the gas kinetic temperature to an intermediate value, $T = 10^{4.5}$~K, for all $U$ and both densities.
A density of $n_{{\rm e}}=1$~cm$^{-3}$ is used to represent the vacuum case.
As shown in Figure~\ref{Oxygen}, DR is not suppressed at such low densities.
A density of $n_{{\rm e}}=10^{10}$~cm$^{-3}$ represents an interesting intermediate density.
Figure~\ref{Oxygen} shows that the DR is moderately suppressed at this density.
The density is typical of the broad emission-line regions of AGNs \citep{Korista:1997} and is a low-to-intermediate density environment in terms of the CRM.
This density is low enough that the CRM effects are significant but not dominant, so a modified two-level approximation should apply.

Suppression of the recombination coefficients will cause the ionization to increase in the two-level limit.
A corrected two-level approximation might then reproduce the intermediate-density rise in the ionization shown in Figures~10~and~11 of C17.
For these densities, the CRM effects are not yet large and the two-level approximation, with modified recombination coefficients, is a reasonable approximation.
At very high densities, where CRM effects are severe, the gas ionization goes over to the Saha--Boltzmann limit and decreases as density increases.
It would be unrealistic to hope that simple corrections to the two-level approximation could recover this limit.

Figures~\ref{Ionization} and Sets~\ref{Appendix:figCollGreen} and ~\ref{Appendix:figPhotGreen} of Appendix~\ref{Appendix:CloudyExamples} show results for fractional abundance when suppression of DR is applied to collisionally ionized and photoionized gas.
The upper panel shows ionization fractions and the dimensionless ratio $n({\rm ion})/n({\rm element})$.
The series of peaks corresponds to successively higher stages of ionization reaching an abundance peak at a particular temperature or ionization parameter.
In the electron collisional case, the temperature of this peak is determined mainly by the ionization potential, the details of the collisional ionization and recombination rate coefficients, and the density suppression of the latter.
In the photoionization case, the peak is sensitive to both the ionization parameter and the shape of the incident radiation field, in addition to the photoionization cross section and recombination rate coefficients.
The lower panel shows the ratio of the ionization fractions for the two densities to make the changes in the ionization easier to see.
Predictions change by approximately a factor of two for O, although other elements can have an order of magnitude change, as Figures~\ref{Appendix:figCollGreen} and ~\ref{Appendix:figPhotGreen} of Appendix~\ref{Appendix:CloudyExamples} show.
The changes are largest for the intermediate ionization stages of O, reflecting the suppression factors shown in Figure~\ref{Oxygen}.
The general trend for the other elements is for the changes to be largest for lower ionization stages and tend to decrease with increasing charge, as suggested by the $q^{-7}$ dependence discussed in \cite{burgsum}.
The conclusion is that suppression can be large, and tends to be greatest for lower ionization stages, but there is considerable scatter introduced by the details of the atomic structure.

\subsection{Photoionization models of AGN broad emission-line regions}
Cloudy includes a large test suite that allows for autonomous testing of the code's predictions.
This includes a number of models of the ``BLR," the broad emission-line line region of a quasar (AGN3).
Because of their great luminosity, spectra of very-high-redshift quasars can be used to measure the chemical evolution of the universe and the growth of black holes at the centers of galaxies across cosmic time.
The BLR is photoionized, as shown by correlations between changes in the continuum and emission lines, and has densities ranging from $10^9$ to $10^{14}$~cm$^{-3}$, densities where suppression of DR is expected to be significant, as originally pointed out by \cite{davidson}.
The Cloudy test suite includes many BLR models and here we will focus on a subset similar to those discussed in the figures in \citep{Korista:1997}.

A photoionization model is parameterized by the shape of the incident ionizing radiation field or SED, the cloud density and column density, its chemical composition, and either the ionization parameter or flux of ionizing photons striking the cloud's surface.
We use the SED and composition given by \citep{Korista:1997} and consider different densities and radiation field intensities.
Table~\ref{TableG1} shows the impact of suppressed DR on predicted line intensities for a number of different BLR models.
The first column gives the identification of various strong UV emission lines.
The remaining columns are for different BLR model parameters.
Each model has a hydrogen density $n({\rm H})$ [cm$^{-3}$] and flux of ionizing photons $\Phi({\rm H})$ [cm$^{-2}$ s$^{-1}$] indicated in the first row as a log.
Cloudy includes a user-adjustable option to set the suppression factors to unity.
Otherwise the suppression factor appropriate for the density and temperature at each point in the cloud is used.
The remainder of the table gives the ratio of the predicted line intensities with, and without, suppression of DR.
\begin{deluxetable}{lccccccccc}
%\tabletypesize{\tiny}
\tabletypesize{\small}
\tablecaption{Ratio of BLR line intensities computed with and without DR suppression.
\label{TableG1}}
\tablewidth{0pt}
\tablehead{
\colhead{Line$\diagdown$Model} & \colhead{ 9, 18 } & \colhead{ 9, 20 } & \colhead{ 11, 20 }  & \colhead{ 12, 19 } & \colhead{ 13, 18 } & \colhead{ 13, 22 } & \colhead{ 14, 18 } & \colhead{ 14, 20 } & \colhead{ 14, 22 }
}
\startdata
\ion{O}{6}~1034  & 0.98 & 0.85 & 0.72 & --   & --   & 0.68 & --   & --   & 0.50 \\
\ion{N}{5}~1240  & 0.98 & 1.21 & 0.78 & --   & --   & 0.80 & --   & 0.70 & 0.56 \\
\ion{H}{1}~1216  & 1.00 & 0.99 & 1.03 & 1.00 & 1.00 & 0.98 & 1.00 & 1.00 & 1.03 \\
\ion{Si}{4}~1397 & 1.02 & 1.05 & 1.02 & 0.80 & 1.00 & 0.95 & --   & 0.92 & 0.83 \\
\ion{C}{4}~1549  & 0.99 & 1.04 & 0.96 & 0.79 & --   & 0.99 & --   & 0.85 & 0.88 \\
\ion{O}{3}]~1666 &0.99  & 2.33 & 1.15 & 0.94 & 1.01 & 1.13 & 1.05 & 1.02 & 0.98 \\
\ion{Al}{3}]~1860&1.00  & 1.15 & 1.04 & 0.73 & 0.92 & 0.97 & 0.93 & 0.82 & 0.83 \\
\ion{C}{3}]~1909 &1.00  & 1.66 & 1.27 & 0.92 & 0.95 & 1.37 & 0.99 & 0.90 & 1.02 \\
\hline
\hline\enddata
\end{deluxetable}

The ionization parameter $U$ is proportional to the ratio of $\Phi({\rm H})$ to $n({\rm H})$, so at a given density it increases as $\Phi({\rm H})$ increases.
High-ionization species such as \ion{C}{4}, \ion{N}{5}, or \ion{O}{6} are not present at low $U$ and the table has no entry for these lines.
The table shows that line intensities generally change by less than a factor of two.
The changes in the intensities are the result of a complex interplay between temperature, ionization, and density, and simple trends are not obvious.
As the flux of ionizing photons increases, the temperature of the gas also tends to increase, making DR more important, but the ionization also increases, with DR suppression becoming less important (the $q^{-7}$ effect shown in Figure~\ref{Oxygen}).
The net effect depends on all of these details.
We stress that the DR suppression factors are highly uncertain, so the changes listed in Table~\ref{TableG1} are only an indication of the types of changes that might occur if a true CRM were done.
This is a high priority for future development.

\section{Summary}
We report on revised and improved Paper~I DR suppression factors, which are to be used as a preliminary test of the extent the finite densities will likely have on the effective DR rate coefficients.
The first group of revisions eliminates potential numerical instabilities that arise in Cloudy simulations and/or modeling that use them.
These instabilities are a consequence of assumptions introduced in Paper~I for the five lowest isoelectronic sequences, and on finer numerical grids, may manifest themselves as temperature and density discontinuities.
The second group of revisions extends the applicability of the suppression factor model to isoelectronic sequences for which secondary autoionization plays an important role.
Improvements are mainly in the reproducibility of collisional-radiative data \citep{summersRAL}, in particular the better prediction of activation densities that mark the onset of suppression of zero-density DR rate coefficients.
As such, the present results are to be used with care outside the Cloudy program, especially if applied to zero-density DR rate coefficients obtained by neglecting the effects of secondary autoionization, where care should be taken to select the appropriate expression for the suppression factor, as discussed in Sec.~\ref{Sec:Suppress}.
Despite the approximations, we stress the importance of density effects on DR processes in astrophysical plasmas and the need for detailed collisional-radiative calculations.

\section{Acknowledgments}

T.W.G. was supported in part by NASA (NNX11AF32G and NNX17AD41G).
K.T.K. was supported in part by NASA (NNX11AF32G).
M.C. acknowledges support from NASA Program number HST-AR-14556.001-A through a grant from the Space Telescope Science Institute.
G.J.F. acknowledges support by NSF (1108928; and 1109061), NASA (10-ATP10-0053, 10-ADAP10-0073, and NNX12AH73G), and STScI (HST-AR-12125.01, GO-12560, and HST-GO-12309).
F.G. acknowledges support by NSF (1412155).
P.v.H. was supported by the Belgian Federal Science Policy Office under contract No.~BR/143/A2/BRASS.
N.R.B. acknowledges support by STFC (ST/J000892/1).

%===============================%
%===% SECTION: BIBLIOGRAPHY  %===%
%===============================%
\bibliographystyle{aasjournal}

\begin{thebibliography}{}
\expandafter\ifx\csname natexlab\endcsname\relax\def\natexlab#1{#1}\fi
\providecommand{\url}[1]{\href{#1}{#1}}
\providecommand{\dodoi}[1]{doi:~\href{http://doi.org/#1}{\nolinkurl{#1}}}
\providecommand{\doeprint}[1]{\href{http://ascl.net/#1}{\nolinkurl{http://ascl.net/#1}}}
\providecommand{\doarXiv}[1]{\href{https://arxiv.org/abs/#1}{\nolinkurl{https://arxiv.org/abs/#1}}}

\bibitem[{{Badnell} {et~al.}(2015){Badnell}, {Ferland}, {Gorczyca},
  {Nikoli\'c}, \& {Wagle}}]{Badnell:2015}
{Badnell}, N.~R., {Ferland}, G.~J., {Gorczyca}, T.~W., {Nikoli\'c}, D., \&
  {Wagle}, G.~A. 2015, \apj, 804, 10, \dodoi{10.1088/0004-637X/804/2/100}

\bibitem[{{Badnell} {et~al.}(2003){Badnell}, {O'Mullane}, {Summers}, {Altun},
  {Bautista}, {Colgan}, {Gorczyca}, {Mitnik}, {Pindzola}, \&
  {Zatsarinny}}]{drproject}
{Badnell}, N.~R., {O'Mullane}, M.~G., {Summers}, H.~P., {et~al.} 2003, \aap,
  406, 1151, \dodoi{10.1051/0004-6361:20030816}

\bibitem[{{Bates} {et~al.}(1962){Bates}, {Kingston}, \& {McWhirter}}]{bates}
{Bates}, D.~R., {Kingston}, A.~E., \& {McWhirter}, R.~W.~P. 1962,
  Proc.~R.~Soc.~Lond.~A, 267, 297, \dodoi{10.1098/rspa.1962.0101}

\bibitem[{{Blaha}(1972)}]{Blaha:1972}
{Blaha}, M. 1972, Astrophysical Letters, 10, 179

\bibitem[{{Burgess} \& {Summers}(1969)}]{burgsum}
{Burgess}, A., \& {Summers}, H.~P. 1969, \apj, 157, 1007,
  \dodoi{10.1086/150131}

\bibitem[{{Davidson}(1975)}]{davidson}
{Davidson}, K. 1975, \apj, 195, 285, \dodoi{10.1086/153328}

\bibitem[{{Ferland} {et~al.}(2017){Ferland}, {Chatzikos}, {Guzm\'{a}n},
  {Lykins}, {van Hoof}, {Williams}, {Abel}, {Badnell}, {Keenan}, {Porter}, \&
  {Stancil}}]{cloudy17}
{Ferland}, G.~J., {Chatzikos}, M., {Guzm\'{a}n}, F., {et~al.} 2017, Revista
  Mexicana de Astronomía y Astrofísica, 53, 385,
  \dodoi{http://www.astroscu.unam.mx/rmaa/RMxAA..53-2/PDF/RMxAA..53-2_gferland.pdf}

\bibitem[{{Hubeny} \& {Mihalas}(2014)}]{Hubeny:2014}
{Hubeny}, I., \& {Mihalas}, D. 2014, Theory of Stellar Atmospheres An
  Introduction to Astrophysical Non-equilibrium Quantitative Spectroscopic
  Analysis, Princeton Series in Astrophysics (41 William Street, Princeton, New
  Jersey 08540: Princeton University Press)

\bibitem[{{Kaur} {et~al.}(2017){Kaur}, {Gorczyca}, \& {Badnell}}]{Kaur:2017}
{Kaur}, J., {Gorczyca}, T.~W., \& {Badnell}, N.~R. 2017, \aap, 610, A41,
  \dodoi{10.1051/0004-6361/201731243}

\bibitem[{{Korista} {et~al.}(1997){Korista}, {Baldwin}, {Ferland}, \&
  {Verner}}]{Korista:1997}
{Korista}, K., {Baldwin}, J., {Ferland}, G., \& {Verner}, D. 1997, \apjs, 108,
  401, \dodoi{10.1086/312966}

\bibitem[{{Nikoli\'c} {et~al.}(2013){Nikoli\'c}, {Gorczyca}, {Korista},
  {Ferland}, \& {Badnell}}]{Nikolic:2013}
{Nikoli\'c}, D., {Gorczyca}, T.~W., {Korista}, K.~T., {Ferland}, G.~J., \&
  {Badnell}, N.~R. 2013, \apj, 768, 1, \dodoi{10.1088/0004-637X/768/1/82}

\bibitem[{{Osterbrock} \& {Ferland}(2006)}]{Osterbrock:2006}
{Osterbrock}, D.~E., \& {Ferland}, G.~J. 2006, Astrophysics of Gaseous Nebulae
  and Active Galactic Nuclei, 2nd edn. (Sausalito, CA: University Science
  Books)

\bibitem[{{Ralchenko}(2016)}]{Ralchenko:2016}
{Ralchenko}, Y. 2016, Springer Series on Atomic, Optical, and Plasma Physics,
  Vol.~90, Validation and Verification of Collisional-Radiative Models, ed.
  Y.~{Ralchenko} (Springer International Publishing), 181--208

\bibitem[{{Ralchenko} {et~al.}(2011){Ralchenko}, {Kramida}, {Reader}, \&
  {NIST~ASD~Team}}]{nist}
{Ralchenko}, Y., {Kramida}, A.~E., {Reader}, J., \& {NIST~ASD~Team}. 2011,
  National Institute of Standards and Technology

\bibitem[{{Summers}(1974 \& 1979)}]{summersRAL}
{Summers}, H.~P. 1974 \& 1979, Appleton Laboratory Internal Memorandum IM367 \&
  re-issued with improvements as AL-R-5

\bibitem[{{Summers}(2004)}]{OPEN-ADAS}
---. 2004, {The ADAS User Manual, version 2.6}, \dodoi{http://www.adas.ac.uk}

\bibitem[{{Summers} \& {Hooper}(1983)}]{sumhoop}
{Summers}, H.~P., \& {Hooper}, M.~B. 1983, Plasma Physics, 25, 1311,
  \dodoi{10.1088/0032-1028/25/12/303}

\bibitem[{{Young}(2018)}]{Young:2018}
{Young}, P.~R. 2018, \apj, 855, 15, \dodoi{10.3847/1538-4357/aaab48}

\end{thebibliography}
%%%\bibliographystyle{aa}

% comment out the next line if you want APPENDIX materials to be included
%\end{document}
% =============================================
%
% APPENDIX MATERIALS
%
% =============================================

\clearpage
\newpage

%\renewcommand{\normalsize}{\fontsize{10}{12}\selectfont}
%\renewcommand{\thefigure}{A-\arabic{figure}}
%\setcounter{figure}{0}
%\renewcommand{\thetable}{A-\arabic{table}}
%\setcounter{table}{0}
%\renewcommand{\thepage}{SM-\arabic{page}}
%\setcounter{page}{1}
%\renewcommand{\thesection}{SM-\arabic{section}}
%\setcounter{section}{0}

%===================================%
%===% APPENDIX: ACCURACY%===%
%===================================%
\appendix
\section{Visualization of approximations to $\psi^N(q,T)$}
\label{Appendix:Psi}
\begin{figure}[!hbtp]
\begin{tabular}{lll}
\includegraphics[width=2.2in]{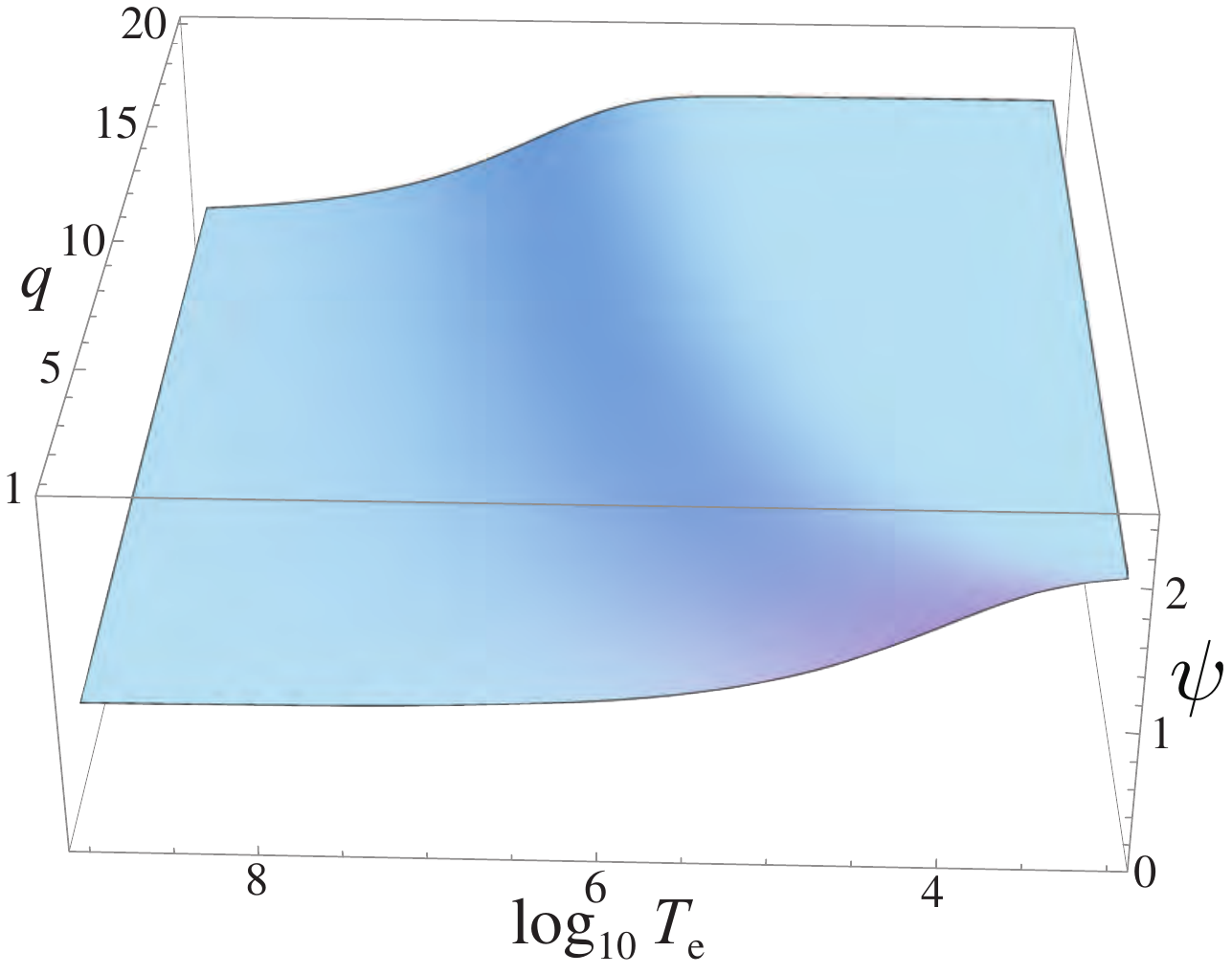}
&
\includegraphics[width=2.2in]{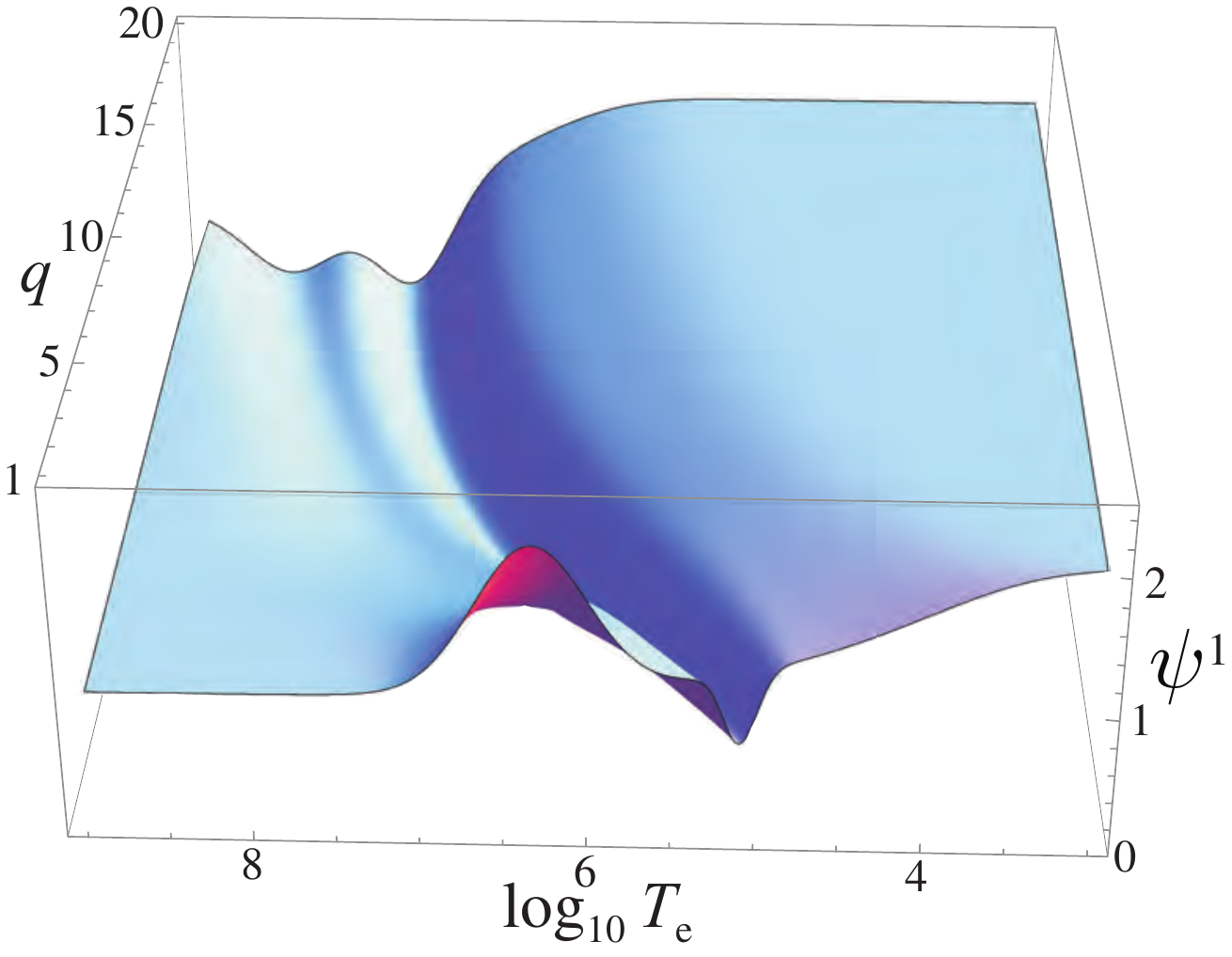}
&
\includegraphics[width=2.2in]{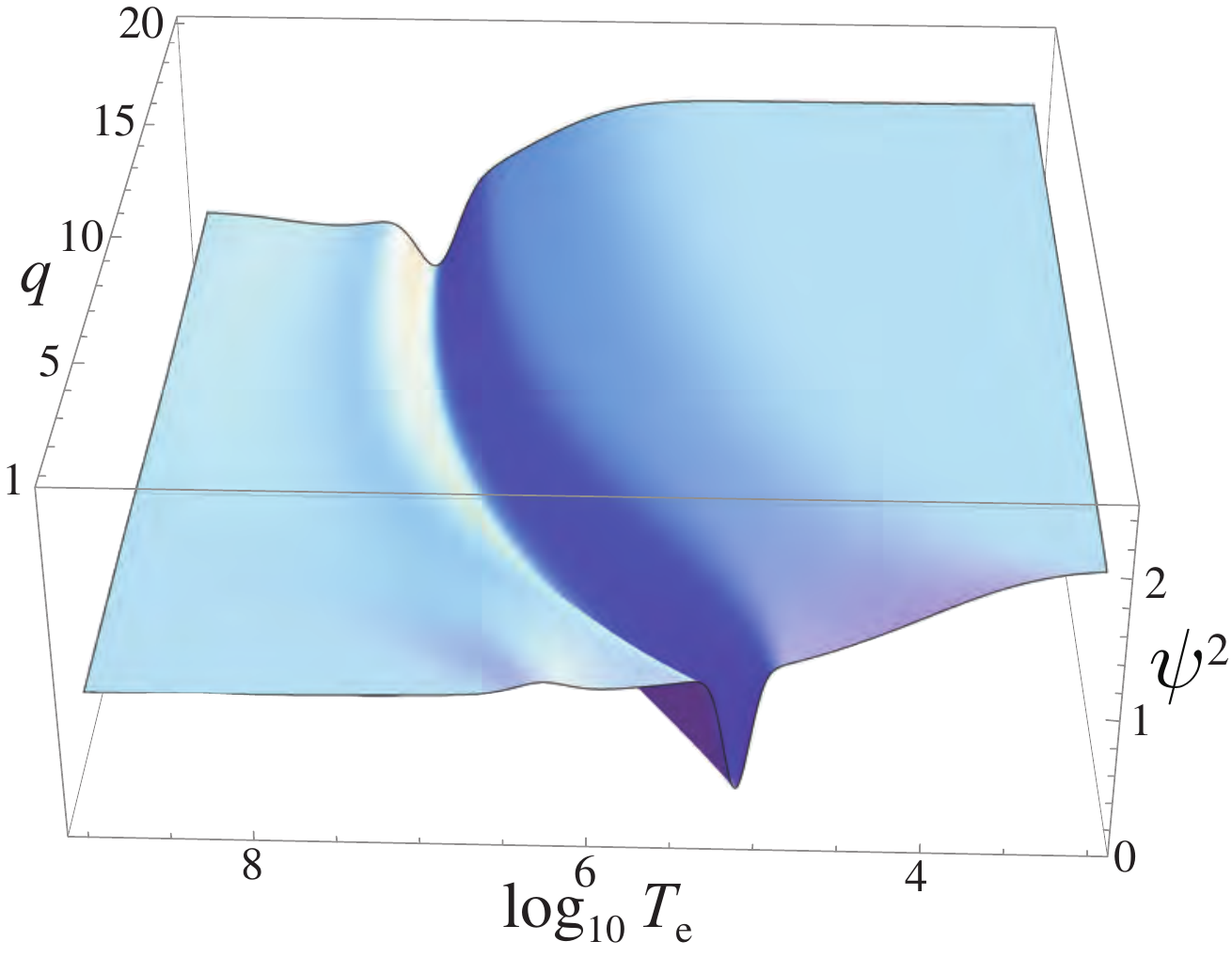}
\\
\includegraphics[width=2.2in]{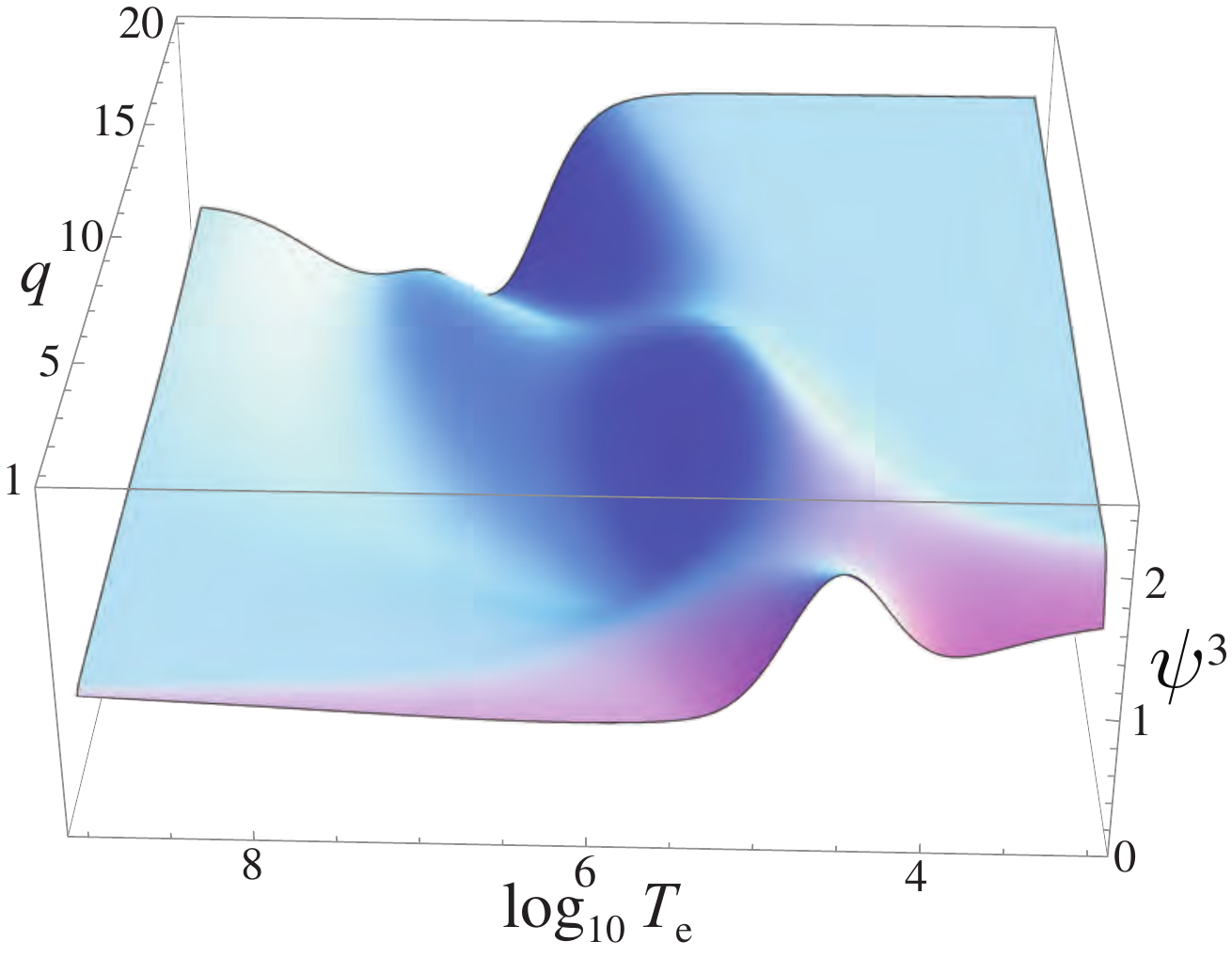}
&
\includegraphics[width=2.2in]{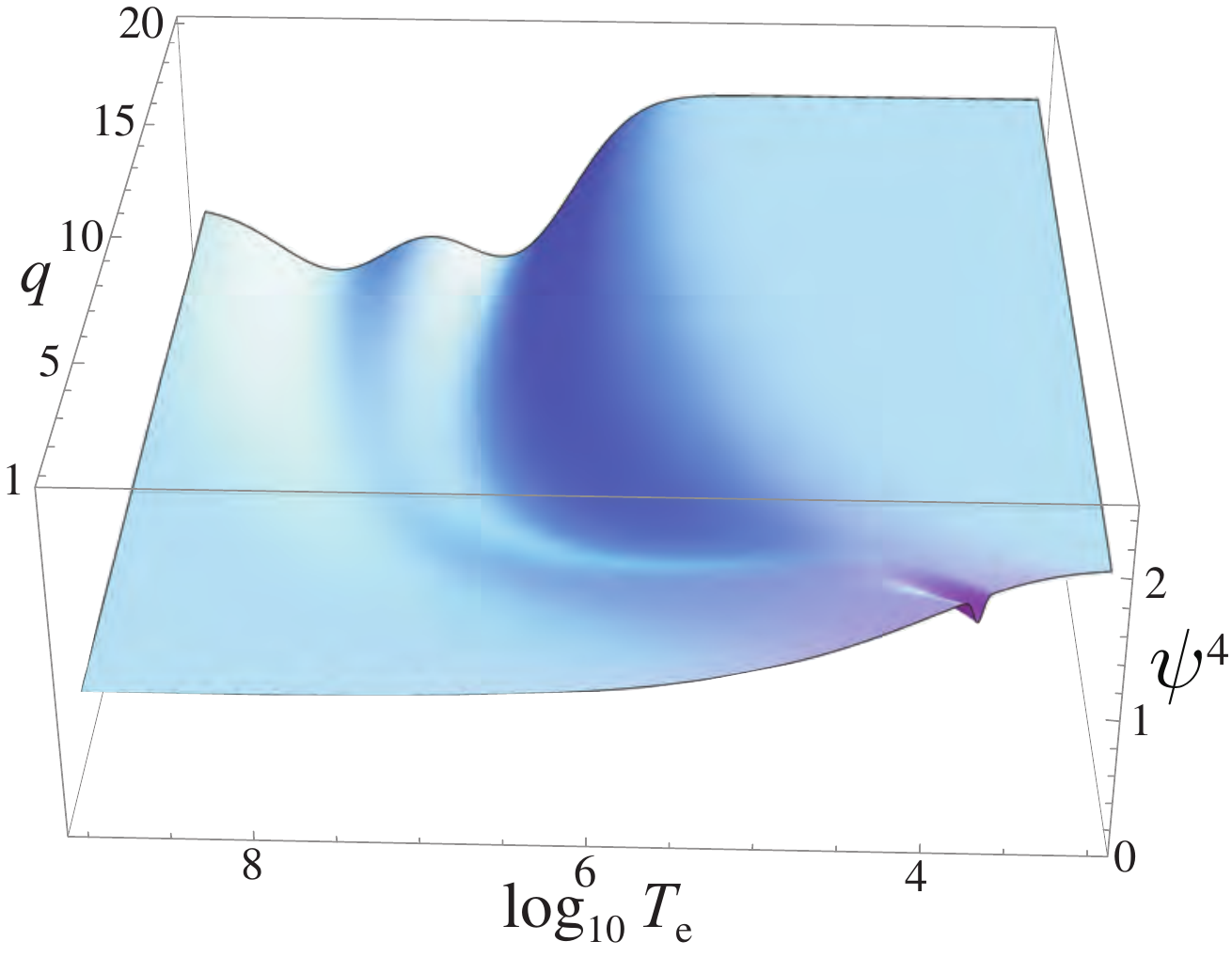}
&
\includegraphics[width=2.2in]{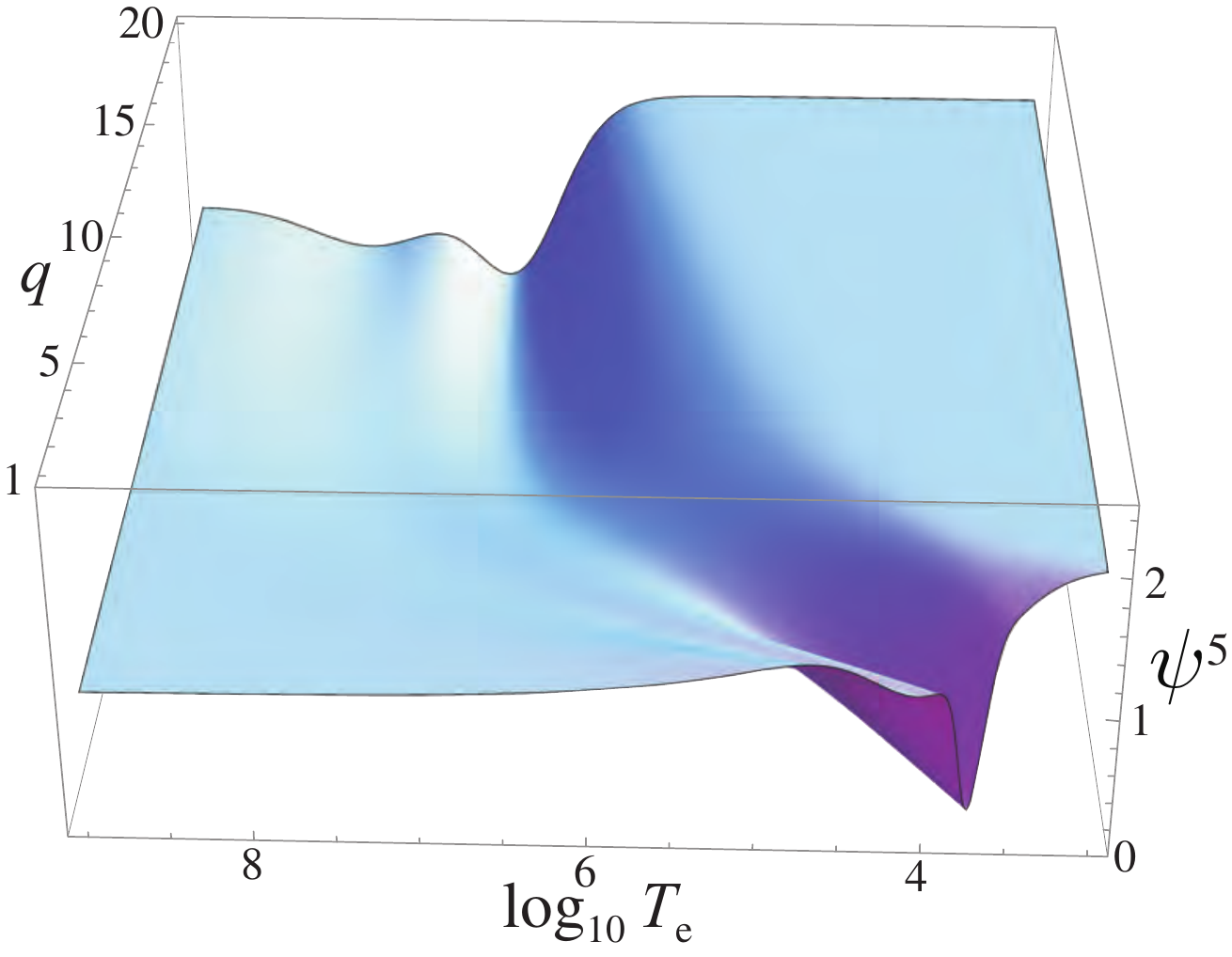}
\\
\includegraphics[width=2.2in]{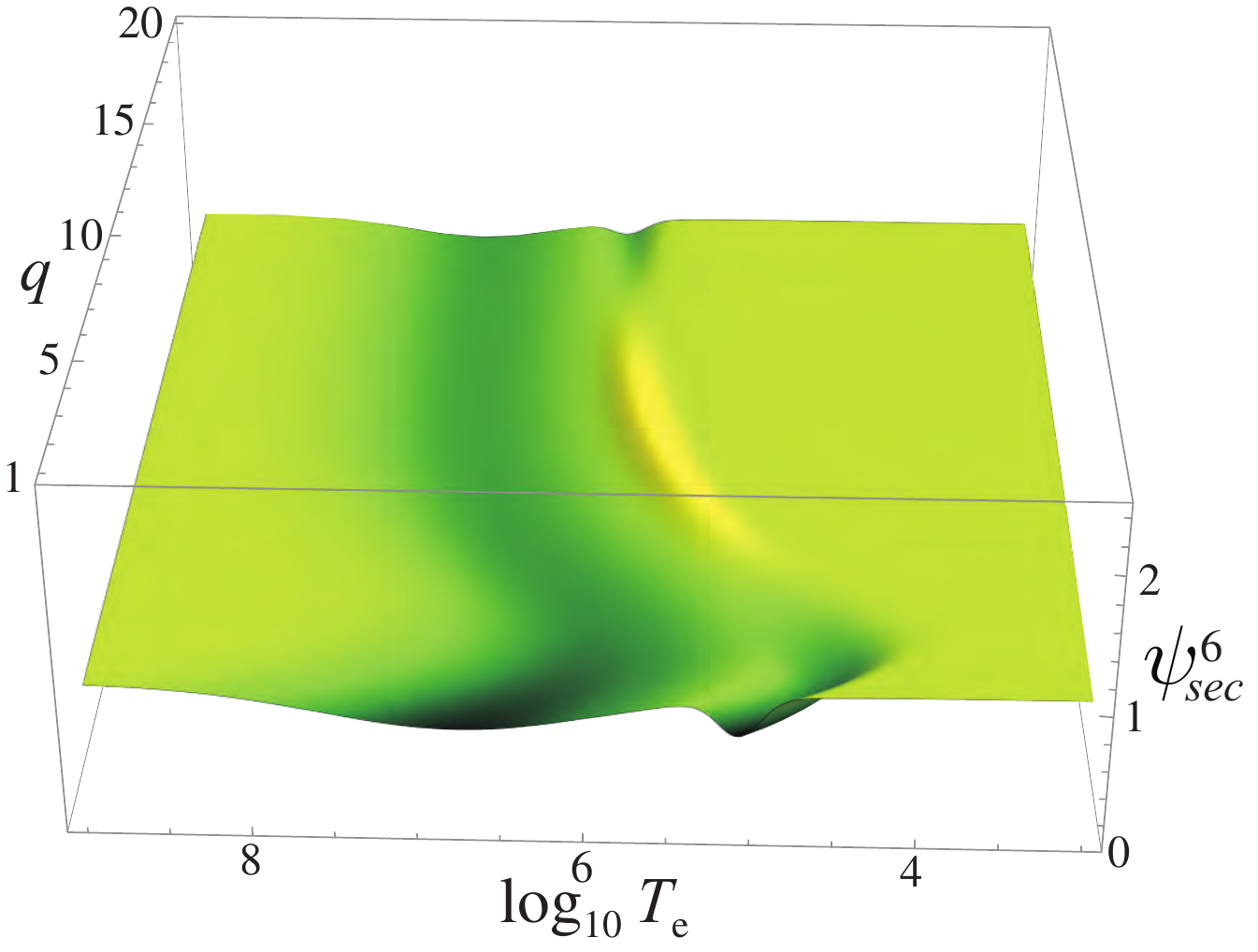}
&
\includegraphics[width=2.2in]{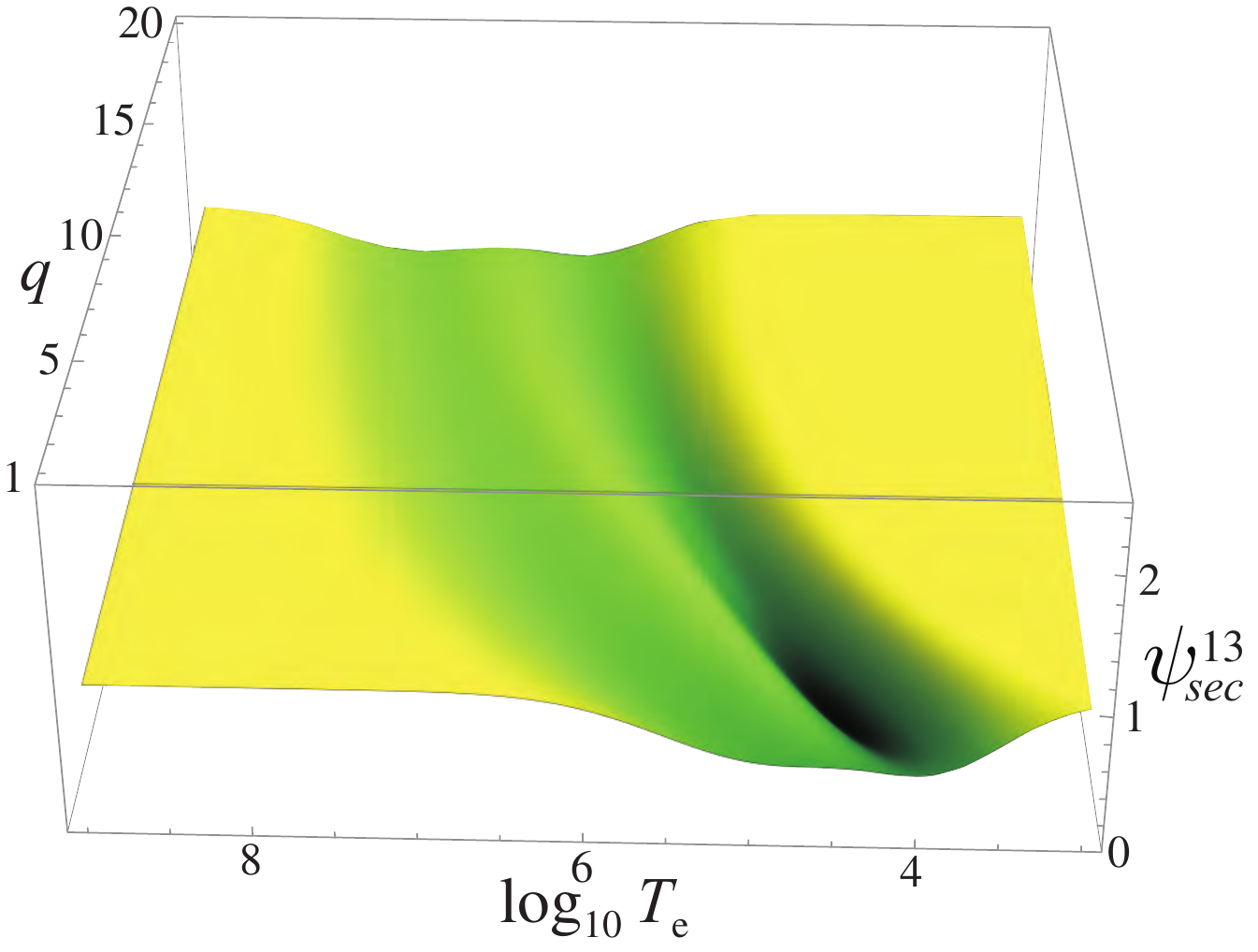}
&
\includegraphics[width=2.2in]{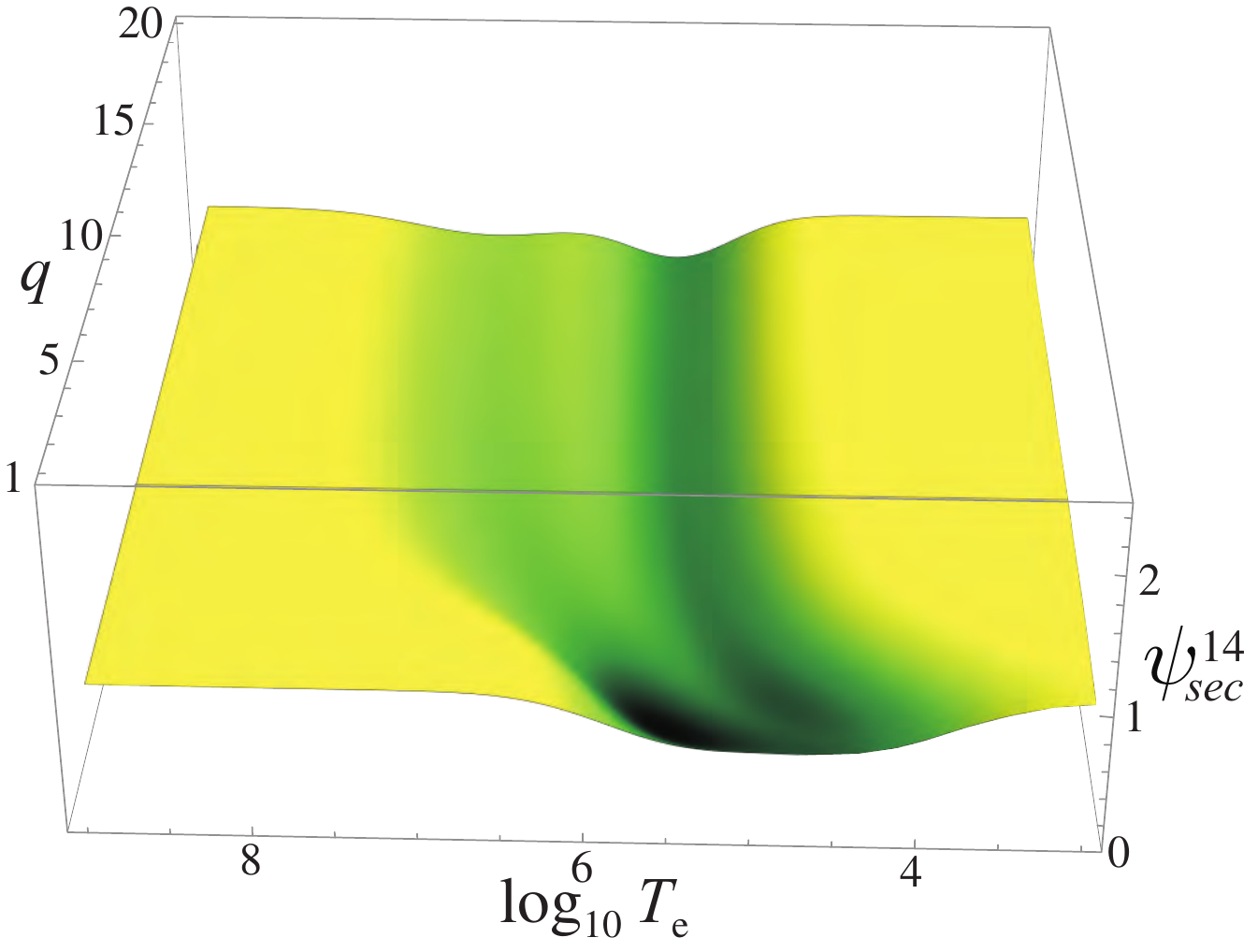}
\end{tabular} \caption{\label{Appendix:figpsi}
Illustration of adjustment factors given in Table~\ref{TablePI}: (top left) ``simplified" $\psi$ and (top/middle) ``detailed" $\psi^N(q,T)$ factors given in Eq.~(\ref{eqFactor}); (bottom) ``secondary autoionization" $\psi^{N}_{sec}(q,T)$ given in Eq.~(\ref{eqFactorSec}).
}
\end{figure}

Figure~\ref{Appendix:figpsi} shows a subset of the model's dimensionless functions $\psi^N(q,T)$ implemented in this study for several isoelectronic sequences. 
These functions modulate the onset of finite-density effects on suppression of DR for each ion species  by changing the characteristic activation densities in plasmas of varying temperatures.
Visual presentation of dimensionless functions $\psi^N(q,T)$ is additionally supported by listing their numerical values in Table~\ref{TablePsiCheck}.

\begin{deluxetable}{c|ccccc|cccc}
%\tabletypesize{\tiny}
\tablecaption{Values of ``detailed" $\psi^N(q,T)$ and ``secondary autoionization" $\psi^{N}_{sec}(q,T)$ given at specified $N$, $q$, and $T$ for checking computer code.
\label{TablePsiCheck}}
\tablewidth{0pt}
\tablehead{
\colhead{$\log_{10}T_e$}  &
\colhead{$\psi^1(7,T)$}&
\colhead{$\psi^2(6,T)$}&
\colhead{$\psi^3(5,T)$}&
\colhead{$\psi^4(4,T)$}&
\colhead{$\psi^5(3,T)$}&
\colhead{$\psi^{5}_{sec}(3,T)$}&
\colhead{$\psi^{6}_{sec}(2,T)$}&
\colhead{$\psi^{13}_{sec}(3,T)$}&
\colhead{$\psi^{14}_{sec}(2,T)$}}
\startdata %done
4.0& 1.99997  & 1.99985 & 2.00606  & 1.9964   & 1.97356 & 0.996451 & 0.99851  & 0.665474 & 0.735488 \\
4.5& 1.99604  & 1.9904  & 2.236    & 1.94375  & 1.49331 & 0.945171 & 0.845034 & 0.352422 & 0.592158 \\
5.0& 1.94122  & 1.90513 & 2.49967  & 1.70856  & 1.01333 & 0.976266 & 0.934901 & 0.594999 & 0.580803 \\
5.5& 1.75124  & 1.68622 & 1.5975   & 1.35216  & 1.11084 & 0.999676 & 0.871361 & 0.556274 & 0.509677 \\
6.0& 1.39431  & 1.1391  & 0.994997 & 1.14642  & 1.09561 & 0.921143 & 0.774866 & 0.659487 & 0.749839 \\
6.5& 0.547837 & 0.91716 & 0.914818 & 0.994872 & 1.10643 & 0.999988 & 0.753717 & 0.822447 & 0.973149 \\
\hline
\enddata
\end{deluxetable}

%===================================%
%===% APPENDIX: ACCURACY%===%
%===================================%
\section{Accuracy of approximations to $\psi^N(q,T)$}
\label{Appendix:Acuracy}

\begin{figure}[!hbtp]
\begin{tabular}{ll}
\includegraphics[width=3.5in]{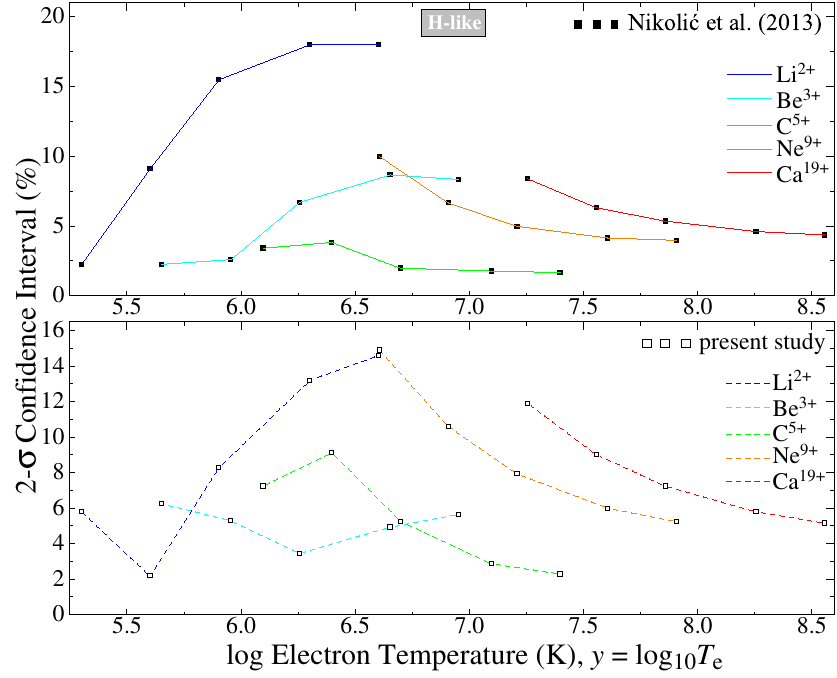}
&
\includegraphics[width=3.5in]{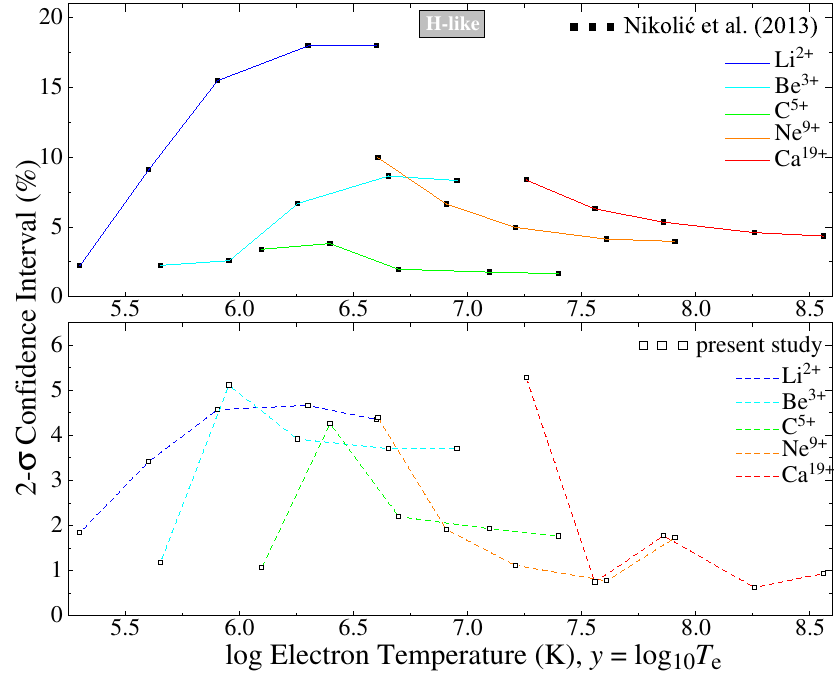}
\\
\includegraphics[width=3.5in]{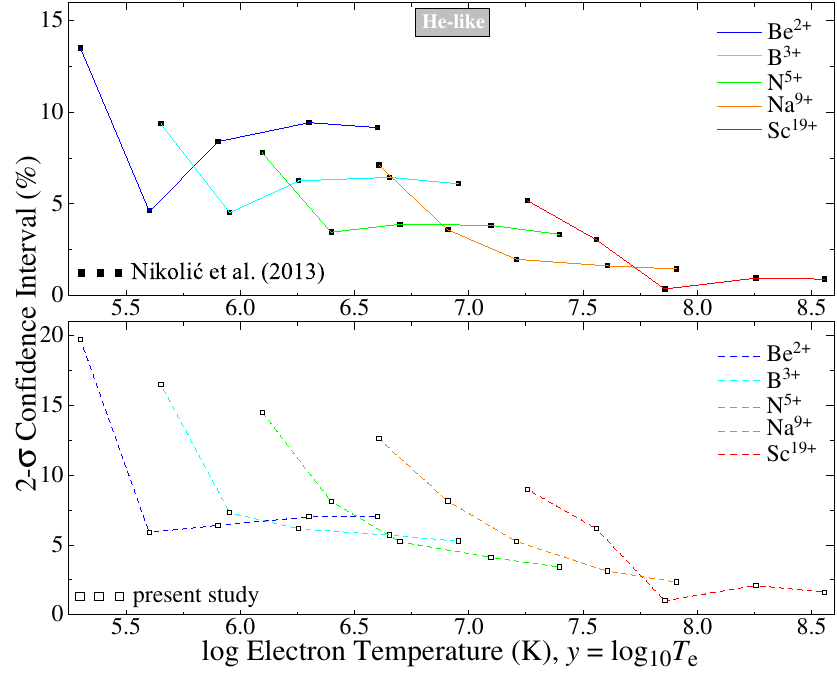}
&
\includegraphics[width=3.5in]{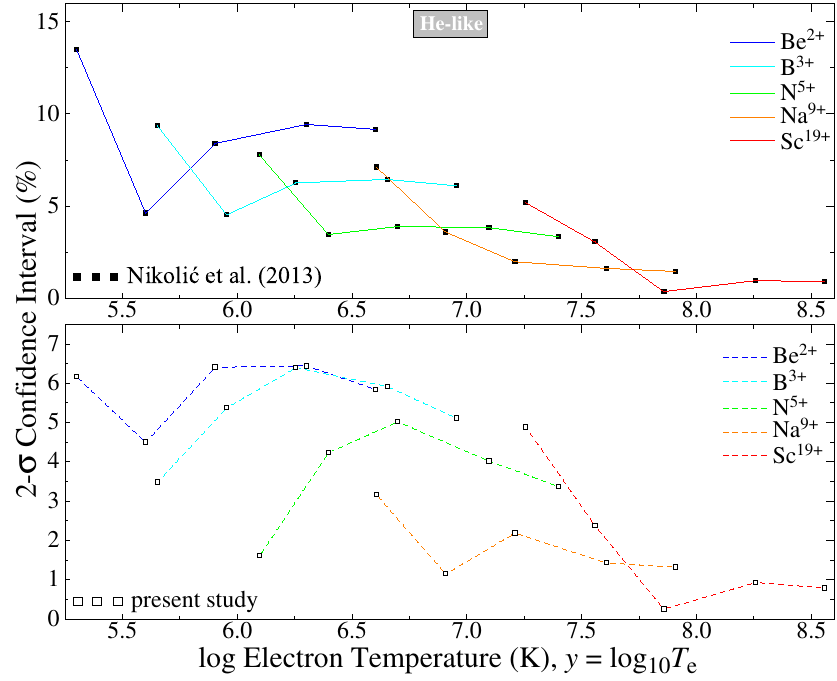}
\\
\includegraphics[width=3.5in]{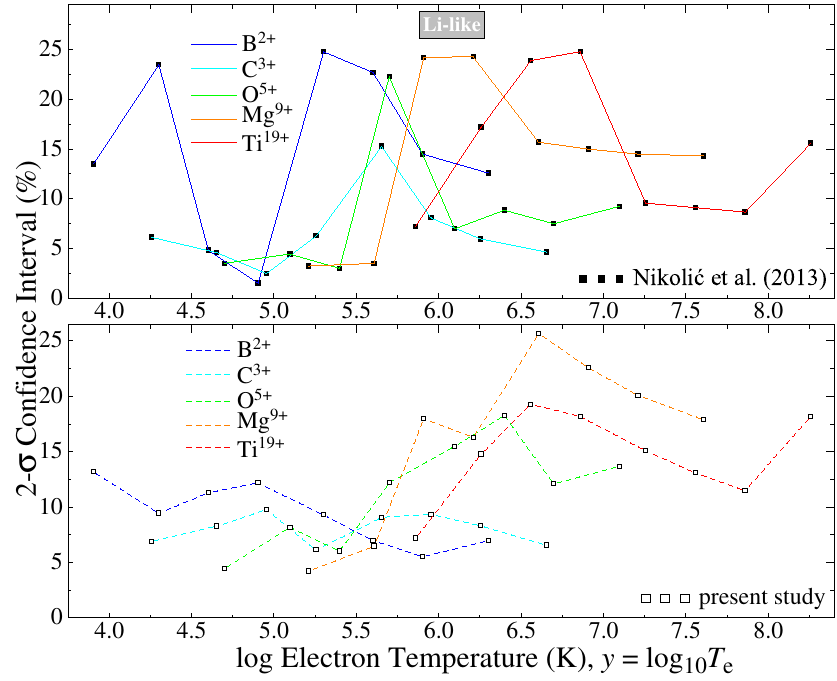}
&
\includegraphics[width=3.5in]{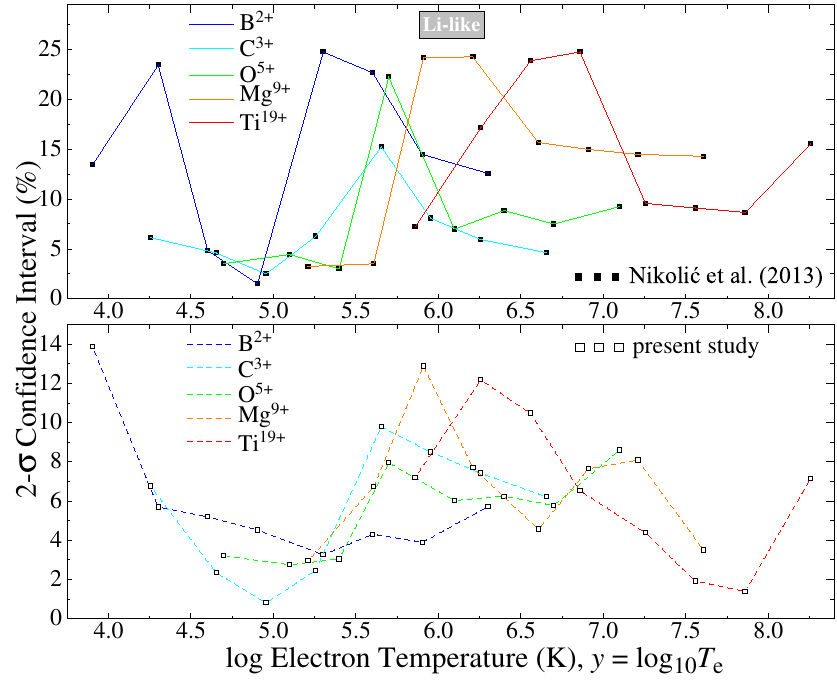}
\end{tabular}
\caption{\label{Appendix:figsurveyOldSmoothGreen}
Estimated accuracy for suppression factors for several charge states as a function of electron temperature when different activation log densities are used. See the text for details.
}
\end{figure}

Figure~\ref{Appendix:figsurveyOldSmoothGreen} illustrates the 2-$\sigma$ (95.4\%) confidence levels from reproducing the suppression factors of \citet{summersRAL} for several charge states as a function of electron temperature.
Higher electron number densities, for which the three-body recombination becomes a dominant process at low $\theta$ values, are excluded from 2$\sigma$ estimates.
The top panels illustrate the accuracies for select ions using the methodology of Paper~I, with the adjustment factor $A^{mod,old}(N)$ given in Eq.~(\ref{eqanold}) and Table~\ref{TableAN}.
The bottom panels are corresponding accuracies from the present study: the left column shows results for ``simplified" $\psi$ and the right column shows results for ``detailed" $\psi^N(q,T)$ adjustment factors, both given in Eq.~(\ref{eqFactor}) and Table~\ref{TablePI}.
In general, when compared to the accuracies of Paper~I, the use of a ``simplified" $\psi$ adjustment factor maintains or slightly improves the accuracy for suppression factors for a wide range of temperatures.
Most importantly, it removes the discontinuity in suppression factors at low temperatures as introduced in Paper~I for isolectronic sequences below C-like.
When activation log densities $x_a(T;q,N)$ given in Eq.~\ref{eqxanew} are evaluated using the ``detailed" $\psi^N(q,T)$ adjustment factors, the overall accuracy improves to better than 14~\%.
\clearpage
\newpage
\FloatBarrier
\clearpage
\newpage

\renewcommand{\thefigure}{SM-\arabic{figure} (Cont.)}
\addtocounter{figure}{-1}
\begin{figure}[!hbtp]
\begin{tabular}{ll}
\includegraphics[width=3.5in]{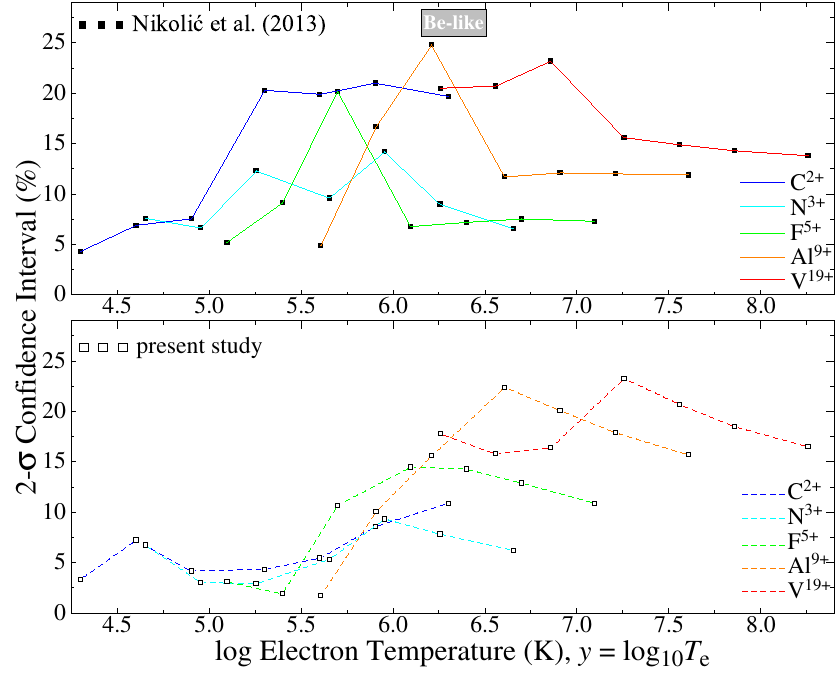}
&
\includegraphics[width=3.5in]{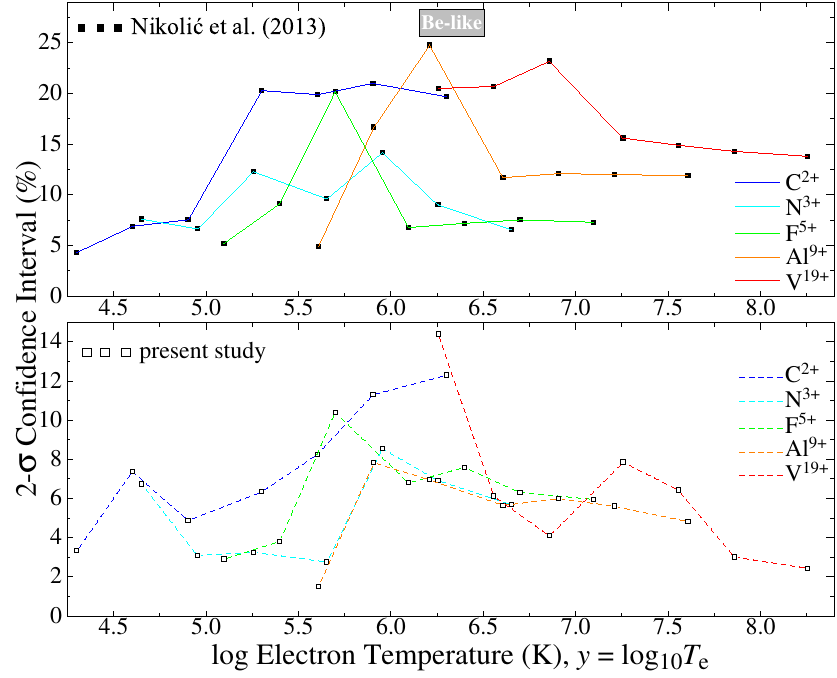}
\\
\includegraphics[width=3.5in]{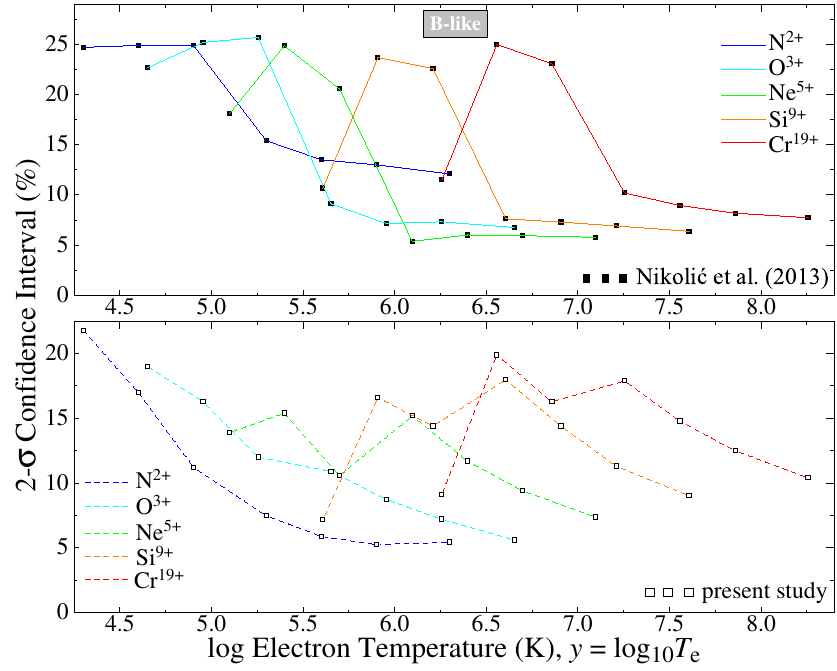}
&
\includegraphics[width=3.5in]{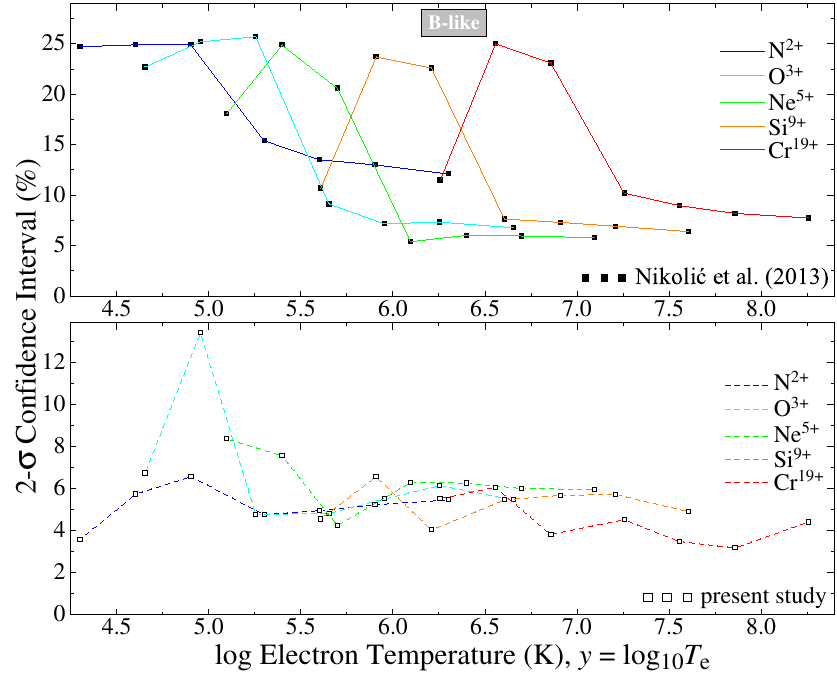}
\end{tabular}
\caption{ \label{Appendix:figsurveyOldSmoothGreenCont2}
}
\end{figure}
\renewcommand{\thefigure}{SM-\arabic{figure}}
\clearpage
\newpage

%===================================%
%===% APPENDIX: ENERGIES%===%
%===================================%
\section{Excitation energies $\epsilon_N(q)$}
\label{Appendix:ExcitationEnergies}
With respect to Paper~I, we update Table~\ref{Appendix:TableEnergies} with the ion-core excitation energy for an Si-like S$^{2+}$ ion to include the results of \citet{Badnell:2015}.

%\vspace*{-15cm}
%\begin{rotatetable}
\begin{deluxetable}{lcrrrrrr}
%\tabletypesize{\tiny}
\tablecaption{\label{Appendix:TableEnergies}
Fitting coefficients for the excitation energies $\epsilon_N(q)=\sum_{j=0}^{5} p_{N,j} \left(\frac{q}{10}\right)^j$, in eV (see Eq.~\ref{eqepsilon}). Note that numbers in square brackets denote powers of 10.}
\tablewidth{0pt}
\tablehead{
\colhead{Sequence} & \colhead{$N$} & \colhead{$p_{N,0}$} & \colhead{$p_{N,1}$} & \colhead{$p_{N,2}$} & \colhead{$p_{N,3}$} & \colhead{$p_{N,4}$} & \colhead{$p_{N,5}$}
}
\startdata
Li-like & 3  & 1.963[+0] &  2.030[+1] & -9.710[-1] &  8.545[-1] & 1.355[-1] &  2.401[-2] \\
Be-like & 4  & 5.789[+0] &  3.408[+1] &  1.517[+0] & -1.212[+0] & 7.756[-1] & -4.100[-3] \\
~N-like & 7  & 1.137[+1] &  3.622[+1] &  7.084[+0] & -5.168[+0] & 2.451[+0] & -1.696[-1] \\
Na-like & 11 & 2.248[+0] &  2.228[+1] & -1.123[+0] &  9.027[-1] & -3.860[-2] &  1.468[-2] \\
Mg-like & 12 & 2.745[+0] &  1.919[+1] & -5.432[-1] &  7.868[-1] & -4.249[-2] &  1.357[-2] \\
~P-like & 15 & 1.428[+0] &  3.908[+0] &  7.312[-1] & -1.914[+0] & 1.051[+0] & -8.992[-2] \\
\hline
H-, He-, Ne-like  & 1,2,10  & $^{\dag}$ & 0.0  &  0.0 &  0.0 & 0.0 & 0.0  \\
\hline
B-, C-, O-, F-like  & 5,6,8,9  & 0.0 & 0.0  &  0.0 &  0.0 & 0.0 & 0.0  \\
Al-, Si-, S-, Cl-like  & 13,14,16,17 & 0.0$^{\ddag}$ & 0.0  &  0.0 &  0.0 & 0.0 & 0.0  \\
\hline
  & $\ge 18$ & 0.0 & 0.0  &  0.0 &  0.0 & 0.0 & 0.0  \\
\hline
\multicolumn{8}{l}{${}^{\dag}$ {\scriptsize $20\;{\rm erfc}(2(x-x_{a}^{0}))$; }${}^{\ddag}$ {\scriptsize set to 17.6874 for Si-like S$^{2+}$, see \citet{Badnell:2015}. }} \\
\hline
\enddata
\end{deluxetable}
%\end{rotatetable}
%\FloatBarrier
\clearpage
\newpage

\section{Examples of Cloudy 17 Model Applications}
\label{Appendix:CloudyExamples}

\subsection{Elemental Abundances Using $\psi^N(q,T)$ and $\psi^{N}_{sec}(q,T)$ Adjustment Factors }
\begin{figure}
\figurenum{A}
\plotone{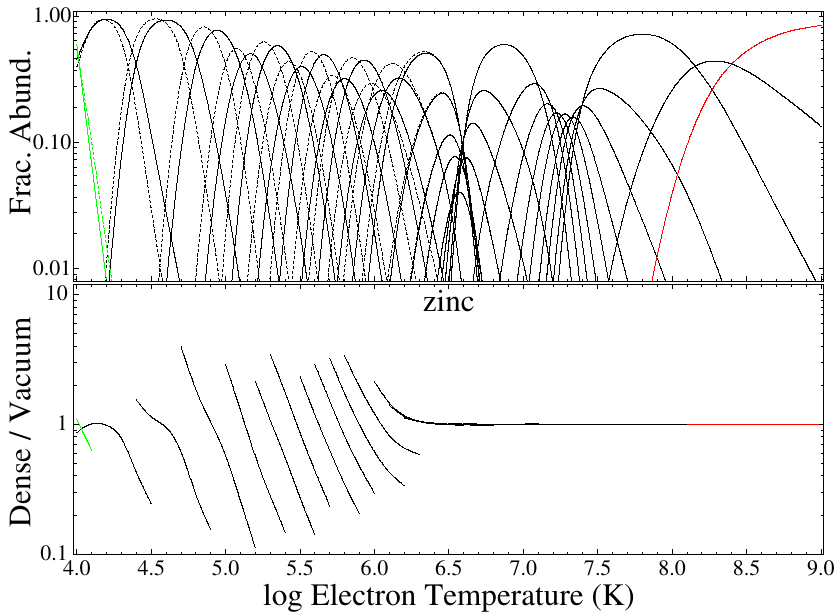}
\caption{\label{Appendix:figCollGreen}
Upper panels: collisional ionization fractional abundance vs. electron temperature for all ionization stages of indicated elements.
Lower panels: ratio of the calculated fractional abundances for the two densities.
The complete figure set (30 images) is available in the online journal.}
\end{figure}

\begin{figure}
\figurenum{B}
\plotone{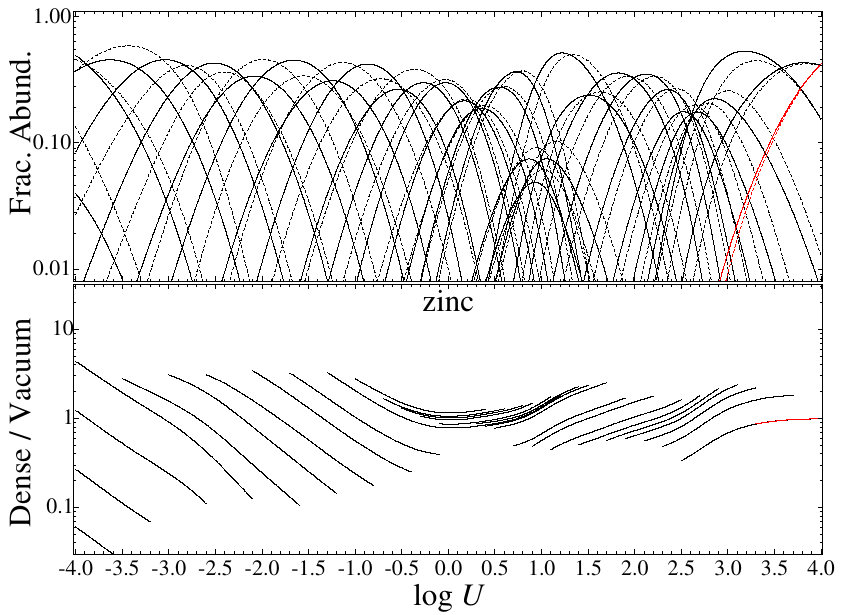}
\caption{\label{Appendix:figPhotGreen}
Upper panels: photoionization fractional abundance vs. the ionization parameter $U$ for all ionization stages of indicated elements and for constant temperature $\log_{10}T_{e}=4.5$.
Lower panels: ratio of the calculated fractional abundances for the two densities.
The complete figure set (30 images) is available in the online journal.}
\end{figure}

Figure~\ref{Appendix:figCollGreen} illustrates finite-density effects on the collisional ionization fractional abundance on all ionization stages of elements up to and including Zn.
All results correspond to the ``detailed" adjustment factor $\psi^N(q,T)$ given in Eq.~(\ref{eqFactor}) and Table~\ref{TablePI}, and where appropriate, to the ``secondary autoionization" $\psi^{N}_{sec}(q,T)$ adjustment factor given in Eq.~(\ref{eqFactorSec}) and Table~\ref{TablePI}.
The solid and dashed curves in the upper panels correspond to electron densities of 1 cm$^{-3}$ and $10^{10}~\rm{cm}^{-3}$, respectively.
From left to right, the curves range from electrically neutral (green) to fully ionized atoms (red).
The lower panels in Figure~\ref{Appendix:figCollGreen} point to the most affected ionization stages by investigating the ratio of the calculated fractional abundances for the two densities.
Similarly, Figure~\ref{Appendix:figPhotGreen} summarizes finite-density effects at constant temperature ($\log_{10}T_{e}=4.5$) on photoionization fractional abundance as a function of ionization parameter $\log_{10}U$.

\end{document}